\documentstyle[jkas,rotating]{article}

\runningauthor{LEE}
\runningtitle{INFRARED SUPERNOVA REMNANTS IN
THE SPITZER GLIMPSE FIELD}
\beginpage{385}
\endpage{414}
\year{2005} \volume{38}
\begin{document}

\title{INFRARED SUPERNOVA REMNANTS IN THE SPITZER GLIMPSE FIELD}

\author{Ho-Gyu Lee}

\address{Astronomy Program, SEES, Seoul National University, Seoul 151-742, Korea \\
{\it E-mail: hglee@astro.snu.ac.kr}}

\address{\normalsize{\it (Received October 5, 2005; Accepted December 7, 2005)}}

\abstract{
We have searched for infrared emission from supernova remnants (SNRs)
included in the Spitzer Galactic Legacy Infrared Mid-Plane Survey Extraordinaire (GLIMPSE)
field.
At the positions of 100 known SNRs, we made 3.6, 4.5, 5.8, and 8.0 \micron\ band images
covering the radio continuum emitting area of each remnant.
In-depth examinations of four band images
based on the radio continuum images of SNRs
result in the identification of sixteen infrared SNRs in the GLIMPSE field.
Eight SNRs show distinct infrared emission in nearly all the four bands,
and the other eight SNRs are visible in more than one band.
We present four band images for all identified SNRs, and RGB-color images
for the first eight SNRs.
These images are the first high resolution ($<$$2''$) images with comparative resolution
of the radio continuum for SNRs detected in the mid-infrared region.
The images typically show filamentary emission along the radio enhanced SNR boundaries.
Most SNRs are well identified in the 4.5 and 5.8 $\mu$m bands.
We give a brief description of the infrared features of the identified SNRs.
}

\keywords{ISM: supernova remnants -- infrared}
\maketitle

\section{INTRODUCTION}

Infrared emission from supernova remnants (SNRs)
is important in the evolution of SNRs; furthermore, it
also delivers unique information pertaining to SNR physics.
Dust grains swept up by SNRs are
heated and disrupted, and emit infrared
radiation that dominate the cooling during most of the
SNR lifetime.
Their infrared emission
contains the most direct informations on
the physical processes associated with dust grains
behind the SNR shocks (Dwek \& Arendt 1992; Draine \& Mckee 1993).
In young SNRs, the thermal infrared emission
may reveal the
composition of newly condensed
dust grains from ejecta, while the non-thermal
infrared synchrotron emission is important for
understanding the acceleration and evolution of high-energy
particles in SNRs.
Ionic and H$_2$ emission lines in infrared emission
serve as a very important
diagnostic tool for understanding the underlying physics
of SNR shocks.
This is particularly true for distant SNRs that suffer large
extinction.
However, despite such usefulness,
the infrared is the least-exploited
waveband in the study of SNRs.

The all-sky survey by the Infrared Astronomical Satellite (IRAS) detected
infrared emission from a significant ($\sim$30\%) fraction of the known SNRs
in the Galaxy (Arendt 1989; Saken et al. 1992).
Thus far, the IRAS survey is the only infrared data set that includes the
entire ensemble of Galactic SNRs.
Although its resolution, of the order of a few-arcminutes,
is useful only to the extent that large SNRs
do not suffer from any confusion effect,
it showed that thermally heated grains are
responsible for the far-infrared emission of SNRs.
In the mid-infrared region,
Kim et al. (2005) reported infrared emission from eight SNRs
using the Galactic plane survey performed by the Midcourse Space Experiment (MSX).
The spectral analysis including the IRAS data suggested that the line contribution
could be significant in the near/mid infrared.
The Infrared Space Observatory (ISO) conducted
spectroscopic and imaging observations of nine SNRs
covering 3--8 \micron\ bands of of the GLIMPSE field.
The majority of them were bright young SNRs
such as Cas A, Tycho, Kepler, and Crab
(e.g., Douvion et al. 1999, 2001a).
Several radiative SNRs interacting with molecular clouds
such as IC 443 and 3C 391 show spatially separated molecular and ionic lines
(Cesarsky et al. 1999; Reach et al. 2002).

GLIMPSE (Benjamin et al. 2003)
is one of the largest surveys performed in the infrared region.
The survey provides fully-sampled, confusion-limited 3.6, 4.5, 5.8 and 8.0 \micron\
Spitzer IRAC data for
the inner two-thirds region of the Galactic plane with
a spatial resolution of $\sim$2$''$.
Most star formation regions, galactic molecular rings, and
more than 4 spiral arms are located in the surveyed area.
The GLIMPSE project focuses on Galactic structure and star formation.
Concurrently, the wide, deep, and unbiased survey provides
a unique opportunity to investigate
diffuse objects, which occupy large areas in the sky.
In this paper, we studied infrared emission
from all SNRs contained in the GLIMPSE Field.
Section 2 explains the data and identification method of
the infrared emission from individual SNRs.
Section 3 reports the identified SNRs and their characteristics.
A brief discussion and summary is included in Section 4.
An independent study to
search for SNRs in the GLIMPSE field (Reach et al. 2005)
exists, and we provide a note on their study at the end of this section.
Finally, we briefly describe the infrared features
of the identified SNRs in the Appendix.

\section{DATA AND IDENTIFICATION OF SNRS}

The GLIMPSE data cover two thirds of the inner Galactic plane
($10^\circ \! < |l| \! < 65^\circ$, $|b|\! < \! 1^\circ$)
in four wavelength bands
centered at 3.6, 4.5, 5.8, and 8.0 \micron\ (Benjamin et al. 2003).
More than 310,000 frames were taken by the
Infrared Array Camera (IRAC) across several days
between March 9 and November 1, 2004.
Each IRAC frame has approximately the same field of view of
5$'$.2 $\times$ 5$'$.2
with a pixel size of $\sim$1$''$.2 (Fazio et al. 2004).
Their point response function varies within the array.
The in-flight mean values of the FWHM are
1$^{\prime \prime}$.66, 1$^{\prime \prime}$.72,
1$^{\prime \prime}$.88, and 1$^{\prime \prime}$.98
in the 3.6, 4.5, 5.8, and 8.0 \micron\ bands, respectively.
The total on-source integration time per position was 4 seconds
after combining the overlapped and revisited frames.
The 5 $\sigma$ sensitivities were estimated
as 0.2 mJy for the two shorter bands and 0.4 mJy for the longer two bands.
The basic calibration and data handling were performed by
the Spitzer Science Center (SSC) Pipeline versions from S9.5.0 to S10.5.0
depending on the observed date.
Further image processing, such as
column pulldown, banding corrections, masking pixel, and removing stray light,
were performed for the image products by the GLIMPSE Pipeline.
We used the GLIMPSE Atlas-Mosaicked Images for each band archived by SSC.
Each mosaicked image was given by a 1$^\circ$.1 $\times$ 0$^\circ$.8 tile
in Galactic coordinates.
Among the 231 Galactic SNRs listed in Green's catalog (2004),
100 SNRs exist in the GLIPMSE field.
We obtained the images of these 100 SNRs in all the four bands.
The images use the equatorial coordinate system
and have a pixel size of 1$^{\prime \prime}$.2.
For large SNRs or for SNRs spreading across the boundaries of the
GLIMPSE Atlas-Mosaicked images, we have coadded adjacent images.
The image processing was made by using the
SWARP software package distributed by
TERAPIX (http://terapix.iap.fr/soft/swarp).
The imaging centers were adopted from Green's catalog,
and the sizes were chosen to cover areas larger than
their respective radio continuum extents:
12$^\prime$ for SNRs smaller than 8$^\prime$,
30$^\prime$ for those between 9$^\prime$ and 20$^\prime$,
and 60$^\prime$ for those larger than 21$^\prime$.

The identification of SNRs requires careful inspection
because they usually appear as faint filaments embedded in a complex field
with confusing sources such as H II regions and planetary nebulae.
For each SNR, we examined all four band images simultaneously
to search for infrared features coincident with the radio continuum emission.
Two radio continuum data sets were used for the comparison,
i.e., NRAO\footnote{The National Radio Astronomy Observatory is
a facility of the National Science Foundation operated
under cooperative agreement by Associated Universities, Inc.}
VLA Sky Survey (Condon et al. 1998) for the northern SNRs and the
MOST\footnote{The MOST is operated by the University of Sydney
with support from the Australian Research Council and
the Science Foundation for Physics within the University of Sydney.}
Supernova Remnant Catalog (Whiteoak \& Green 1996) for the southern SNRs.
The identified SNRs were compared in detail
with high-resolution radio continuum images for confirmation.

\begin{table*}
\begin{sideways}
\scriptsize
\renewcommand\arraystretch{1.3}
\begin{tabular}{lcccc ccccl}
\\ \\ \\ \\ \\
\multicolumn{10}{c}{{\bf Table 1.}~~SNRs Identified in the Spitzer GLIMPSE field} \\
\\
\hline\hline
Name & RA$^a$ & Dec$^a$ & Size$^a$ & Type$^a$
& \multicolumn{4}{c}{IRAS Flux$^b$ (Jy)}
& IRAC Morphology
\\ \cline{2-5} \cline{6-9}
     & (h m s) & ($^\circ$   $^\prime$) & ($^\prime \! \times \! ^\prime$) &
& 12& 25& 60& 100
&    \\
\hline
Distinctly identified SNRs \\ \cline{1-1}
G31.9+0.0 (3C391)    & 18 49 25 & $-$00 55 & 7$\times$5 & S
& 3& 14& 180& 160
& Thick shell broken-out toward southeast
\\
&  &  &  &
& (2)& (7)& (90)& (80)
&
\\
G34.7--0.4 (W44)      & 18 56 00 & +01 22 & 35$\times$27 & C
& 410& 440& 4400& 13000
& Bright thin filaments inside the entire remnant
\\
&  &  &  &
& (100)& (110)& (1100)& (6500)
&
\\
G43.3--0.2 (W49B)     & 19 11 08 & +09 06 & 4$\times$3 & S
& 15&  140& 1100& 2000
& Bright thin filaments inside the entire remnant
\\
&  &  &  &
& (3)& (28)& (220)& (400)
&
\\
G304.6+0.1 (Kes 17)  & 13 05 59 & $-$62 42 & 8 & S
& 17&  24& 160& 440
& Bright thin filaments along the western shell
\\
&  &  &  &
& (12)& (17)& (110)& (310)
&
\\
G311.5--0.3           & 14 05 38 & $-$61 58 & 5 & S
& $<$4& $<$12& $<$190& $<$310
& Almost complete thin filamentary ring
\\
&  &  &  &
&   &   &   &
&
\\
G332.4--0.4 (RCW 103) & 16 17 33 & $-$51 02 & 10 & S
& $<$200& $<$260& $<$3000& $<$5400
& Bright thin filament along the southern shell
\\
&  &  &  &
&   &   &   &
&
\\
G348.5+0.1 (CTB 37A) & 17 14 06 & $-$38 32 & 15 & S
& $<$150& $<$250& $<$2500& $<$8000
& Thick shell broken-out toward southwest
\\
&  &  &  &
&   &   &   &
&
\\
G349.7+0.2           & 17 17 59 & $-$37 26 & 2.5$\times$2 & S
& 12& 48& 370& $<$550
& Very bright filament with bright spot at center
\\
&  &  &  &
& (4)& (16)& (120)&
&
\\
Marginally identified SNRs \\ \cline{1-1}
G11.2--0.3            & 18 11 27 & $-$19 25 & 4 & C
& 44& 189& 1400& 3100
& Thin partial filament at the radio-bright portion
\\
&  &  &  &
& (4)& (19)& (140)& (310)
& of southeastern shell
\\
G21.8--0.6 (Kes 69)   & 18 32 45 & $-$10 08 & 20 & S
& 106& 121& 1370& 6300
& Thin filament at the middle of radio shell
\\
&  &  &  &
& (16)& (18)& (210)& (950)
&
\\
G39.2--0.3 (3C396)    & 19 04 08 & +05 28 & 8$\times$6 & C
& $<$56& $<$82& $<$530& $<$960
& Filamentary emission spread over the western part
\\
&  &  &  &
&   &   &   &
& of remnant
\\
G41.1--0.3 (3C397)    & 19 07 34 & +07 08 & 4.5$\times$2.5 & S
& $<$12& $<$23& $<$224& $<$485
& Very faint filament at the radio enhanced portion
\\
&  &  &  &
&   &   &   &
& of northeastern shell
\\
G54.1+0.3            & 19 30 31 & +18 52 & 1.5 & F?
& $<$21& 18& 26& $<$600
& Diffuse emission at western boundary with extended
\\
&  &  &  &
&   & (6)& (9)&
& tail toward north
\\
G298.6--0.0           & 12 13 41 & $-$62 37 & 12$\times$9 & S
& $<$110& $<$230& $<$1940& $<$3650
& Bright but suspected filament along southern shell
\\
&  &  &  &
&   &   &   &
&
\\
G340.6+0.3           & 16 47 41 & $-$44 34 & 6 & S
& $<$23& $<$32& $<$370& $<$1100
& Thin filament at south overlapped with unrelated
\\
&  &  &  &
&   &   &   &
& confusing emission
\\
G346.6--0.2           & 17 10 19 & $-$40 11 & 8 & S
& $<$160& $<$180& $<$1500& $<$5900
& Thin filament at south overlapped with unrelated
\\
&  &  &  &
&   &   &   &
& confusing emission
\\
\hline
\\
\multicolumn{10}{l}{$^a$Referred from the Green's catalog (2004).} \\
\multicolumn{10}{l}{$^b$The first row is the measured flux and
the second row in parentheses is the uncertainty of measuring flux
(Arendt 1989).} \\
\end{tabular}
\end{sideways}
\end{table*}

\section{RESULTS}

Among the 100 SNRs examined, eight bright SNRs are distinctly identified.
They are listed in Table~1.
The position, size, and type in columns (2)--(5) are quoted from the Green's catalog.
These SNRs show infrared emission in all four bands at the peak position,
and the extent of the emission is comparable to the remnant size.
Figures~1 to 8 show their images.
The gray-scale range varies from the median of the image pixels to
three times of their standard deviations.
This, however, does not hold in the case of two SNRs,
which show a variation of up to
five times for W 49B and seven times for G349.7+0.2.
In each figure, a radio continuum map
is also included at the bottom left
to assist in the verification of the associated infrared emission features.
The position and scale of radio continuum maps are
set to match those of the infrared images.
The radio continuum maps have angular resolutions of
2$^{\prime \prime}$--5$^{\prime \prime}$
except several SNRs in the southern sky and the large SNR W 44.
Thus, in most cases,
thin filamentary features in the infrared can be directly compared
to the corresponding features in radio.
Figures~9 to 16 show RGB-color images of these SNRs created from
8.0 \micron\ (Red), 5.8 \micron\ (Green), and 4.5 \micron\ (Blue) images.
The dynamic range of each color is identical to its gray-scale image.

In addition to the above eight SNRs, we have identified another eight SNRs
that are marginally bright enough to be discernible from the background.
They are also listed in Table~1,
and their images are shown in Figures~17 to 24.
The scale range of the figures are narrower than
(typically half of) those for the distinctly-identified SNRs
to show the associated emission more clearly.
All these SNRs are identified in the 4.5 and 5.8 \micron\ bands
except G54.1+0.3, which is identified only in the 8.0 \micron\ band.
G11.2--0.3 and 3C 396 are also visible in the 8.0 \micron\ band.
G298.6--0.0 has a bright filament visible in all four bands;
however, its association is questionable.

The GLIMPSE images show that the infrared emission of SNRs usually appear
as partial filaments or almost complete thin shells along the SNR boundary.
Kes 17, G311.5--0.3, and RCW 103 are good examples. All marginally-identified
SNRs except G54.1+0.3 belong to this category.
In W 44 and W 49B, the filaments are also seen inside the remnants,
which might be located on the shell in a projected manner.
The partial filaments are found typically at the positions
of enhanced radio emission and are aligned along the radio shell.
However, the coincidence decreases for
the detailed structures.
For example,
the radio-bright western shell has no infrared counterpart in W 44.
In addition, with regard to the fields under investigation,
all the radio-bright SNRs do not have associated infrared emission.
Two SNRs, - 3C 391 and CTB 37A - have rather thick shells,
and they appear to have broken-out shell morphology like radio features.
G349.7+0.2 is unique on account of its very bright infrared emission
in the central area of the remnant.
This remnant is interacting with a molecular cloud at the center
(Lazendic et al. 2005), and the infrared-bright region coincides with
the position of the molecular cloud.
For each SNR, we give a brief description of their infrared features
in the Appendix.

The infrared emission of SNRs are better identified in the 4.5 and 5.8 \micron\ bands
than in the other two bands.
In the 3.6 \micron\ band, the infrared emission from the SNRs is weak.
Concurrently, the emission from stars
makes it difficult to identify the SNR emission.
On the other hand, the 8.0 \micron\ band
is severely affected by
Polycyclic Aromatic Hydrocarbon (PAH) emission at 6.2, 7.7, and 8.6 \micron,
which is dominant in H II regions, planetary nebulae,
and other diffuse Photodissociation Regions (PDRs)
(van Dishoeck 2004).
Therefore, the emission from SNRs is easily confused with
those from other sources.
Hence, most of the identified SNRs
show blue (4.5 \micron) or green filaments (5.8 \micron)
in the RGB images.

\section{DISCUSSION AND SUMMARY}

We have identified 16 (8 distinctly and 8 marginally)
infrared SNRs in the GLIMPSE field. This is
16\% of the 100 radio/X-ray SNRs in the field.
For comparison, previous IRAS studies detected infrared emissions from
22 SNRs out of the 70 (Arendt 1989) or
18 SNRs out of the 75 (Saken et al. 1992),
which were identified
in the GLIMPSE field at the time of the respective studies.
Among the 16 GLIMPSE SNRs, 9 SNRs are detected by IRAS.
IRAS fluxes or limits are listed in columns (6)--(9) of Table~1.
A study was conducted to search for the infrared emission associated with
SNRs from the Midcourse Space Experiment (MSX) data (Kim et al. 2005). MSX
surveyed the whole Galactic plane ($|b| \! < \! 5^\circ$)
in four bands centered at
8.28, 12.13, 14.65, and 21.34 \micron. Kim et al. (2005) generated and
examined the MSX images of over 200 SNRs in the field and identified
8 SNRs, five of which are included in the GLIMPSE field.
These five SNRs,
- G11.2--0.3, W 44, W 49B, G54.1+0.3, and G349.7+0.2 -
are identified in this work.
The higher detection rate of the GLIMPSE survey is
due to its higher sensitivity and angular resolution.
The 0.4 mJy sensitivity of the 8 \micron\ band in the GLIMPSE field
is seven times better than
the 30 mJy sensitivity of the 8.28 \micron\ band of MSX
(Benjamin et al. 2003).
Furthermore, the $18''.3$ resolution of MSX
increases the chance to dilute thin filaments
and confuse them with other sources.
Another large-scale IR survey is the Two Micron All Sky Survey (2MASS),
which surveyed the whole sky at
$J$ (1.25 \micron), $H$ (1.65 \micron), and $K_s$ (2.17 \micron).
No systematic study of SNRs using the 2MASS data exists;
however, four
SNRs (Crab, IC 443, Cas A, and RCW 103) have been reported as visible.
Only RCW 103 exists in the GLIMPSE field.
We checked 2MASS images at the position of
the other GLIMPSE SNRs and were barely able to trace
the emission associated with W 49B and W 44.

In general, the infrared emission from SNRs is composed of
a thermal continuum from shock-heated dust grains, line emission from
shock-heated gas, and synchrotron emission.
Among the 16 GLIMPSE SNRs, only one is Crab-like (or filled-center)
and three are
composite with a central pulsar wind nebulae (PWNe) (see Table~1).
We have not detected any
IR emission corresponding to these synchrotron nebulae.
Except G54.1+0.3, all the 16 SNRs
show shell-like or filamentary IR emission.
If we consider the steep slope of the synchrotron
emission and the low sensitivity of the GLIMPSE data,
it is highly unlikely that the synchrotron
emission contributes to the observed IR flux
even in the shortest IRAC wave band (3.6 \micron).
On the other hand, previous ISO spectroscopic
studies of SNRs showed that ionic lines
[Fe II] 5.340 $\mu$m and [Ar II] 6.985 $\mu$m
in the IRAC bands are particularly strong in radiative shocks
(Arendt 1999; Oliva et al. 1999;
Douvion et al. 1999, 2001a, 2001b; Reach et al. 2002).
For SNRs interacting with molecular clouds, e.g., 3C 391 and IC 443,
shock-excited H$_2$ lines are also found to be strong
(Cesarsky et al. 1999; Reach \& Rho 2000; Reach et al. 2002).
Therefore the ionic lines produced by
the atomic J-type shock (Hartigan et al. 2004)
and/or the H$_2$ lines produced by
the molecular C-type shock (Hollenbach et al. 1989)
can be the main contributors in IRAC images.

Independent of this work,
there exists a study on the
search for SNRs in the GLIMPSE field (Reach et al. 2005).
They also detected the associated infrared emission by visually
inspecting the IRAC images with an overlaid radio map.
18 SNRs were reported to have an associated infrared emission (score~1).
All eight distinct SNRs,
as well as the five of the eight marginally detected SNRs
were included in their list.
The others (G54.1+0.3, G298.6--0.0, G340.6+0.3)
were regarded as SNRs that were not convincingly detected
due to the confusion with unrelated emissions.
For the detected SNRs, the origin of
the infrared emission,
e.g., molecular or ionic shocks or the PAH emission, was discussed
using the relative intensities between the IRAC bands.

The GLIMPSE images of SNRs presented in this paper are the first
high-resolution infrared images for most SNRs. They show that SNRs
appear as almost complete shells, partial filaments, or a
combination of filaments. The infrared emission generally coincides
with the radio continuum, e.g., the infrared emission is identified
with the radio-bright regions, although some displacement is a
feature of the detailed distribution. Most are relatively well
identified in the 4.5 and 5.8 \micron\ bands. The dominant
contribution may originate from the line emission of shocked ambient
gas. We expect that the infrared imaging and spectroscopic
observations using the SST and the future ASTRO-F mission will
extend the understanding of the nature of the infrared emission in
individual sources.
\\

\acknowledgments{ This work is based on archival data obtained
with the Spitzer Space Telescope, which is operated by the Jet
Propulsion Laboratory, California Institute of Technology under a
contract with NASA. I wish to thank the Spitzer GLIMPSE team for
their efforts. This work was supported by the Korea Science and
Engineering Foundation Grant ABRL 3345-20031017. I wish to thank
the Institute of Space and Astronautical Science (ISAS) for
providing their facilities during part of this work. I wish to
thank Bon-Chul Koo for extensive advice on the entire study. I
also wish to thank David Moffett, Stephen Reynolds, Kristy Dyer,
David Green, John Dickel,and Crystal Brogan for prividing the
high-resolution radio continuum images, and Chris Pearson for the
advice to improve the presentation. }

\vspace{2cm}

\begin{center}
{\large \bf APPENDIX. IDENTIFIED SNRS}
\end{center}

G11.2--0.3 -- A faint wispy filament is seen along the SE SNR boundary at 4.5,
5.8, and 8.0 \micron. The $\sim$$20^{\prime \prime}$-long filament centered at
($\rm{18^h11^m31^s.7,-19^\circ27'14''}$) is
particularly noticeable.
The bright emission at
($\rm{18^h11^m19^s.3,-19^\circ26'19''}$) is
not associated with the remnant but with a K1/K2 III star (HD 166422).
\\

G21.8--0.6 (Kes 69) -- A faint $\sim$$6'$-long filament centered at
($\rm{18^h33^m05^s.9,-10^\circ12''53'}$) is seen
along the incomplete SNR shell at 4.5 and 5.8 \micron.
There is a complex, extended emission at
5.8 and 8.0 \micron\ in the NE boundary of the field.
The SE boundary of this structure
appears to form a concave filament that is roughly parallel with the SNR radio
shell. The relation of this emission to the remnant needs to be explored.
\\

G31.9+0.0 (3C391) -- A bright incomplete shell
coincident with the radio structure is seen.
The NW part of the remnant is bright primarily at 5.8 \micron,
while the NE and SW parts are bright at 4.5, 5.8, and 8.0 \micron;
this explains their different
colors in the true-color image. The bright region at
($\rm{18^h49^m23^s.3,-00^\circ57'49''}$)
shows a region where the remnant interacts with
a molecular clump (Reach \& Rho 1999).
\\

G34.7+0.4 (W44) -- A bright filamentary elliptical shell
coincident with the radio shell is seen in all the four
bands. It is most clearly seen at 4.5 \micron,
which yields a blue color
in the true-color image. The eastern edge of the infrared shell is missing
although it is bright in the radio image.
\\

G39.2--0.3 (3C396) -- A faint $\sim$$2^\prime$-long filament
is seen along the SW SNR boundary
where the radio emission is enhanced.
It is most clearly visible at 4.5 \micron. The emission at
($\rm{19^h03^m56^s.2,+05^\circ25'43''}$)
is also be identified in other bands.
\\

G41.1--0.3 (3C397) -- Very faint filaments coincident with
the NE SNR boundary are visible
at 4.5, 5.8, and 8.0 \micron.
The identified filaments are partly affected by the emission
from a bright confusing star located at
($\rm{19^h07^m33^s.6,+07^\circ09'24''}$).
\\

G43.3--0.2 (W49B) -- A bright filamentary shell
matching the radio structure is seen in all the four bands.
The SW portion of the shell is very bright as it is in the radio.
The interior of the remnant is filled with diffuse
emission including several bright filaments running along the NE-SW
direction. A bow-shaped structure bright at 8.0 \micron\ exists just outside
the eastern boundary of the remnant; its association needs to be explored.
\\

G54.1+0.3 --
A partially complete, circular shell is seen mainly at 8.0 \micron. The western
half of the shell is bright and thick, whereas the eastern half is faint and
fuzzy. The remnant is Crab-like in the radio and the shell appears
to surround the remnant.
In addition, a diffuse, extended emission emanate from the
north of the remnant.
\\

G298.6--0.0 --
A $\sim$$5^\prime$-long filament coincident with the southern, radio-bright
SNR shell is seen in all the four bands.
However, the remnant is located in a complex
field, and the possibility of a chance coincidence
cannot be ruled out.
\\

G304.6+0.1 (Kes 17) --
Bright filaments are distinct in all the four bands.
They are distributed in the western part of the remnant
along its boundary.
At 5.8 and 8.0 \micron, the filaments appear to be
connected to the edge of a large cloud
in the NE by a diffuse, extended emission.
No infrared counterpart is detected at the south of the remnant
where the radio emission is stronger than the western shell.
\\

G311.5--0.3 --
An almost complete thin shell with a diameter of $\sim 3^\prime$
is seen in all the four bands.
It is brightest at 4.5 and 5.8 \micron.
The western part of the shell is brighter than the eastern part.
\\

G332.4--0.4 (RCW 103) --
A sharp filament corresponding to the southern part of the remnant is
visible in all the four bands. It is particularly bright at  4.5 \micron.
The bright portion of the filament is at the position
where the radio brightness has a sharp
cutoff in contrast to the other parts of the remnant.
There is also a faint infrared filament corresponding to the
western radio-bright region.
\\

G340.6+0.3 --
A complex filamentary structure superposed on the northern part of the
remnant is visible at 5.8 and 8.0 \micron. However, it is not correlated
with the radio structure of the remnant, and its association needs to
be explored. Some diffuse emission coincident with the
southern SNR shell also exists.
\\

G346.6--0.2 --
A very faint thin filament delineating the SE boundary of the SNR shell is
visible at 4.5 and 5.8 \micron. There exists some faint emission  coincident with
the northern SNR shell; however,
its association is not clear because of
the confusing extended structure to the west of the remnant.
\\

G348.5+0.1 --
A large ($\sim$$10^\prime$) incomplete shell is
seen mainly at 5.8 and 8.0 \micron.
The remnant is composed of a small, bright NE portion
and a large, diffuse SW portion
in the radio, and the infrared shell coincides with the SW portion.
The infrared shell is brighter toward the NE
where the radio structure starts to blow out.
The remnant appears to be connected to a diffuse, extended structure
at the NE.
\\

G349.7+0.2 --
Bright filamentary emission is visible in the interior of the remnant
in all the four bands.
A very bright $\sim$$40^{\prime \prime}$-long,
"v"-shaped filament at the center and a bright
$\sim$$45^{\prime \prime}$-long filament perpendicular
to the former in the south are prominent.
These filaments coincide with the position of the molecular clouds
swept up by the SNR shock (Lazendic et al. 2005).
Another bright filament stemming out from the
eastern end of the "v"-shaped filament is visible along the NE SNR shell.
A long filamentary emission is connected to the south of the remnant at 5.8
and 8.0 \micron. This emission stretches out $\sim$$3^\prime$
toward the east.

\newpage
\begin{figure*}
\vspace{-3cm}
\centerline{\epsfysize=24cm\epsfbox{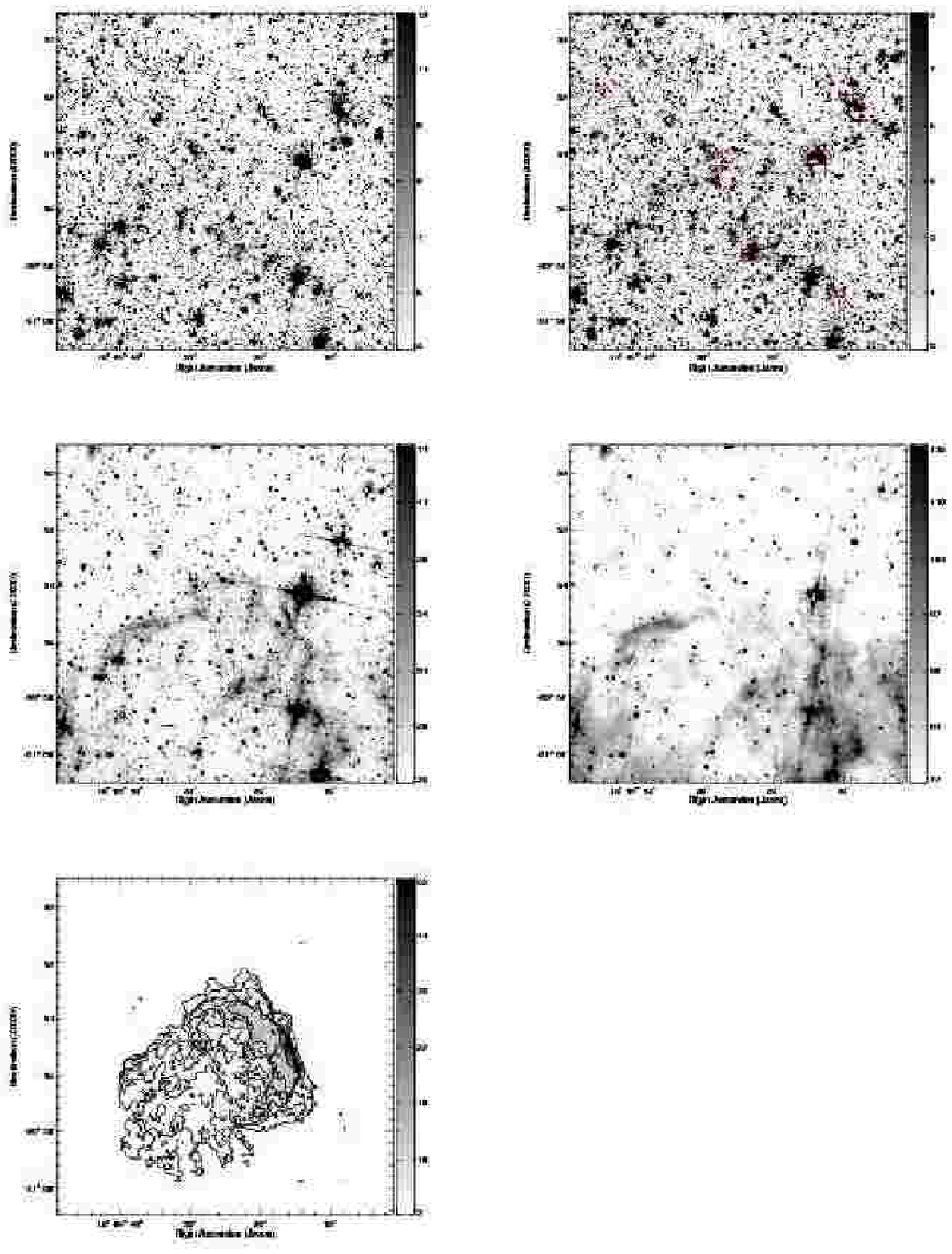}}
{\bf Fig. 1.}---~
IRAC images of G31.9+0.0 (3C 391) at
   3.6 \micron\ (top-left),
   4.5 \micron\ (top-right),
   5.8 \micron\ (middle-left), and
   8.0 \micron\ (middle-right).
   The gray-scale range
   is shown at the left of each image.
   VLA 1.46 GHz continuum image with
   $5^{\prime \prime}.0 \! \times \! 4^{\prime \prime}.6$ beam (Moffett \& Reynolds 1994)
   is also shown at the bottom-left.
   The contour levels are
   1, 2.5, 5, 7.5, 10, 20, and 40 mJy beam$^{-1}$.
\end{figure*}

\newpage
\begin{figure*}
\vspace{-3cm}
\centerline{\epsfysize=24cm\epsfbox{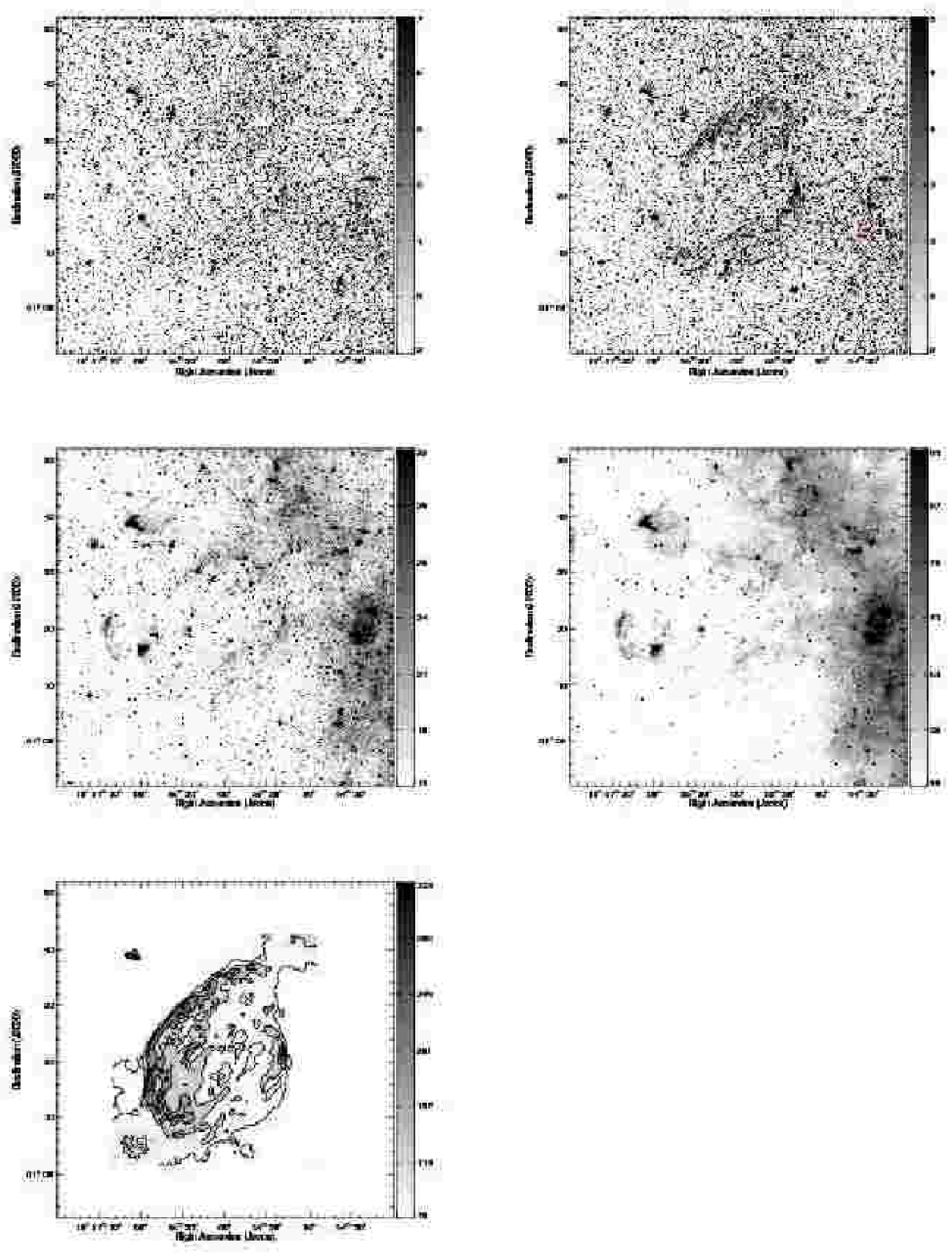}}
{\bf Fig. 2.}---~
IRAC images of G34.7--0.4 (W 44) at
   3.6 \micron\ (top-left),
   4.5 \micron\ (top-right),
   5.8 \micron\ (middle-left), and
   8.0 \micron\ (middle-right).
   The gray-scale range
   is shown at the left of each image.
   VLA 1.44 GHz continuum image with
   $30^{\prime \prime}.0 \! \times \! 30^{\prime \prime}.0$ beam
   is also shown at the bottom-left.
   The contour levels are
   50, 100, 150, and 200 mJy beam$^{-1}$.
\end{figure*}

\newpage
\begin{figure*}
\vspace{-3cm}
\centerline{\epsfysize=24cm\epsfbox{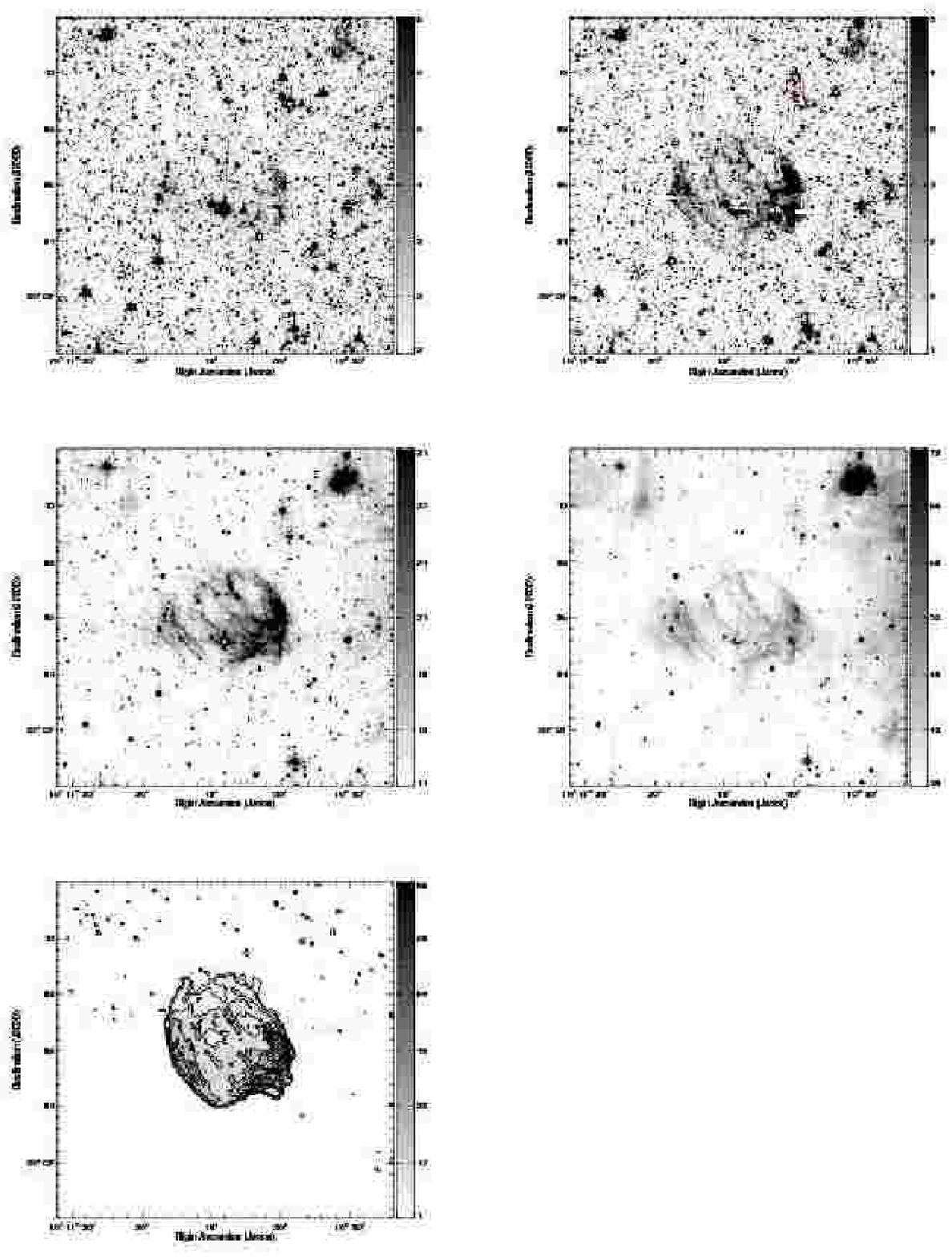}}
{\bf Fig. 3.}---~
IRAC images of G43.3--0.2 (W 49B) at
   3.6 \micron\ (top-left),
   4.5 \micron\ (top-right),
   5.8 \micron\ (middle-left), and
   8.0 \micron\ (middle-right).
   The gray-scale range
   is shown at the left of each image.
   VLA 1.45 GHz continuum image with
   $5^{\prime \prime}.2 \! \times \! 4^{\prime \prime}.8$ beam (Moffett \& Reynolds 1994)
   is also shown at the bottom-left.
   The contour levels are
   3, 5, 10, 20, 30, 40, and 50 mJy beam$^{-1}$.
\end{figure*}

\newpage
\begin{figure*}
\vspace{-3cm}
\centerline{\epsfysize=24cm\epsfbox{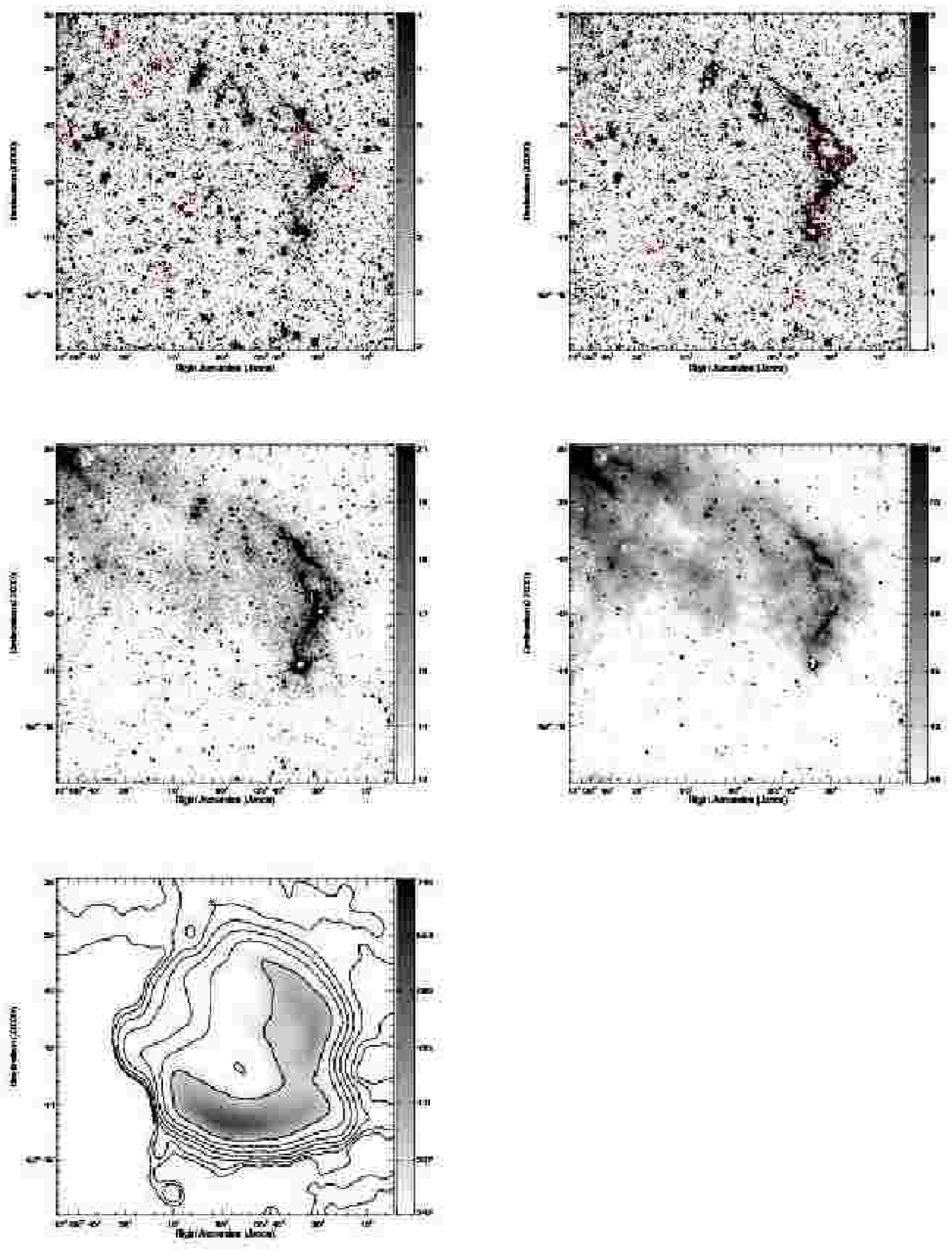}}
{\bf Fig. 4.}---~
IRAC images of G304.6+0.1 (Kes 17) at
   3.6 \micron\ (top-left),
   4.5 \micron\ (top-right),
   5.8 \micron\ (middle-left), and
   8.0 \micron\ (middle-right).
   The gray-scale range
   is shown at the left of each image.
   MOST 0.843 GHz continuum image with
   $43^{\prime \prime} \! \times \! 43^{\prime \prime}$ beam
   is also shown at the bottom-left.
   The contour levels are
   5, 10, 20, 40, 80, 160, and 320 mJy beam$^{-1}$.
\end{figure*}

\newpage
\begin{figure*}
\vspace{-3cm}
\centerline{\epsfysize=24cm\epsfbox{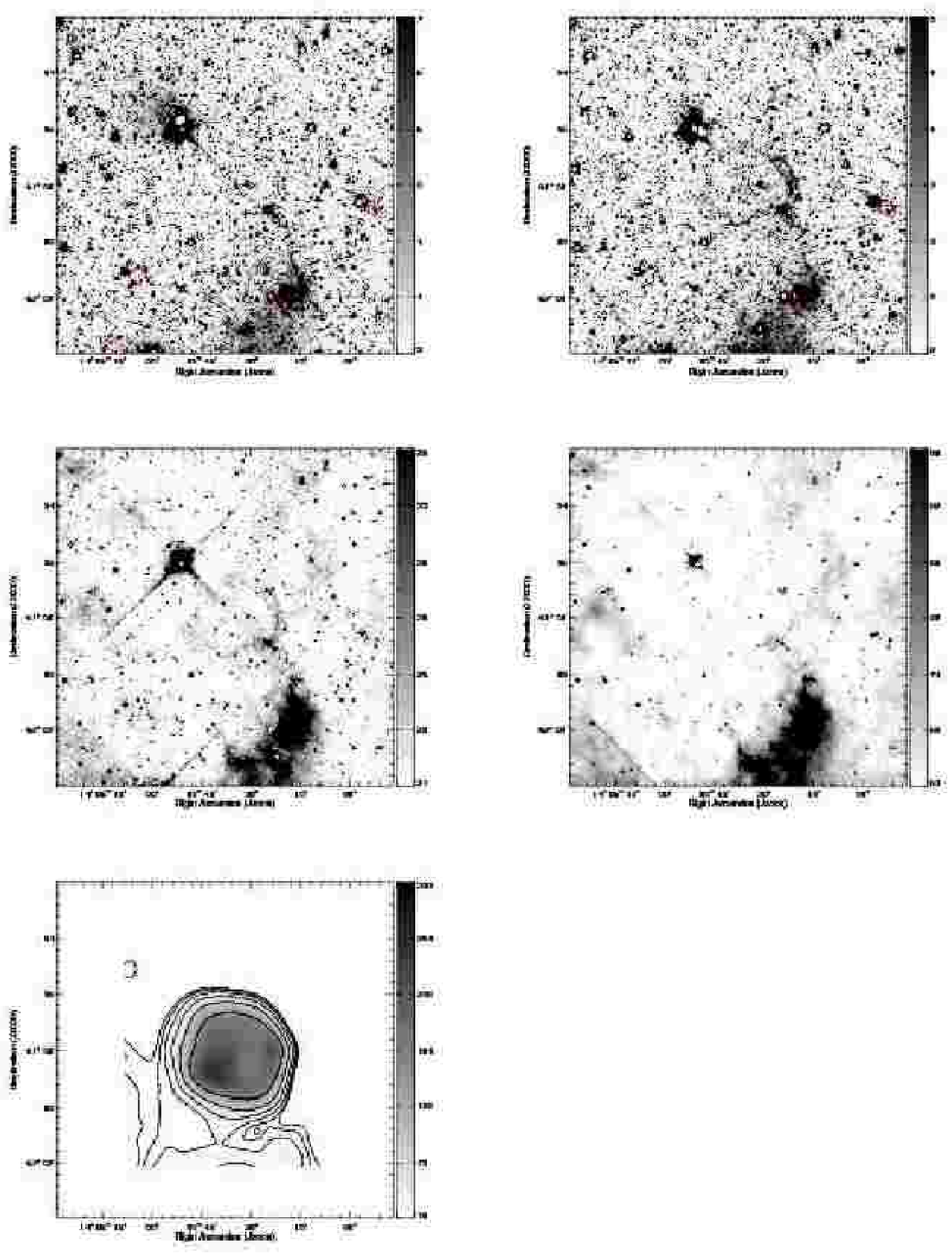}}
{\bf Fig. 5.}---~
IRAC images of G311.5--0.3 at
   3.6 \micron\ (top-left),
   4.5 \micron\ (top-right),
   5.8 \micron\ (middle-left), and
   8.0 \micron\ (middle-right).
   The gray-scale range
   is shown at the left of each image.
   MOST 0.843 GHz continuum image with
   $43^{\prime \prime} \! \times \! 43^{\prime \prime}$ beam
   is also shown at the bottom-left.
   The contour levels are
   5, 10, 20, 40, 80, and 160 mJy beam$^{-1}$.
\end{figure*}

\newpage
\begin{figure*}
\vspace{-3cm}
\centerline{\epsfysize=24cm\epsfbox{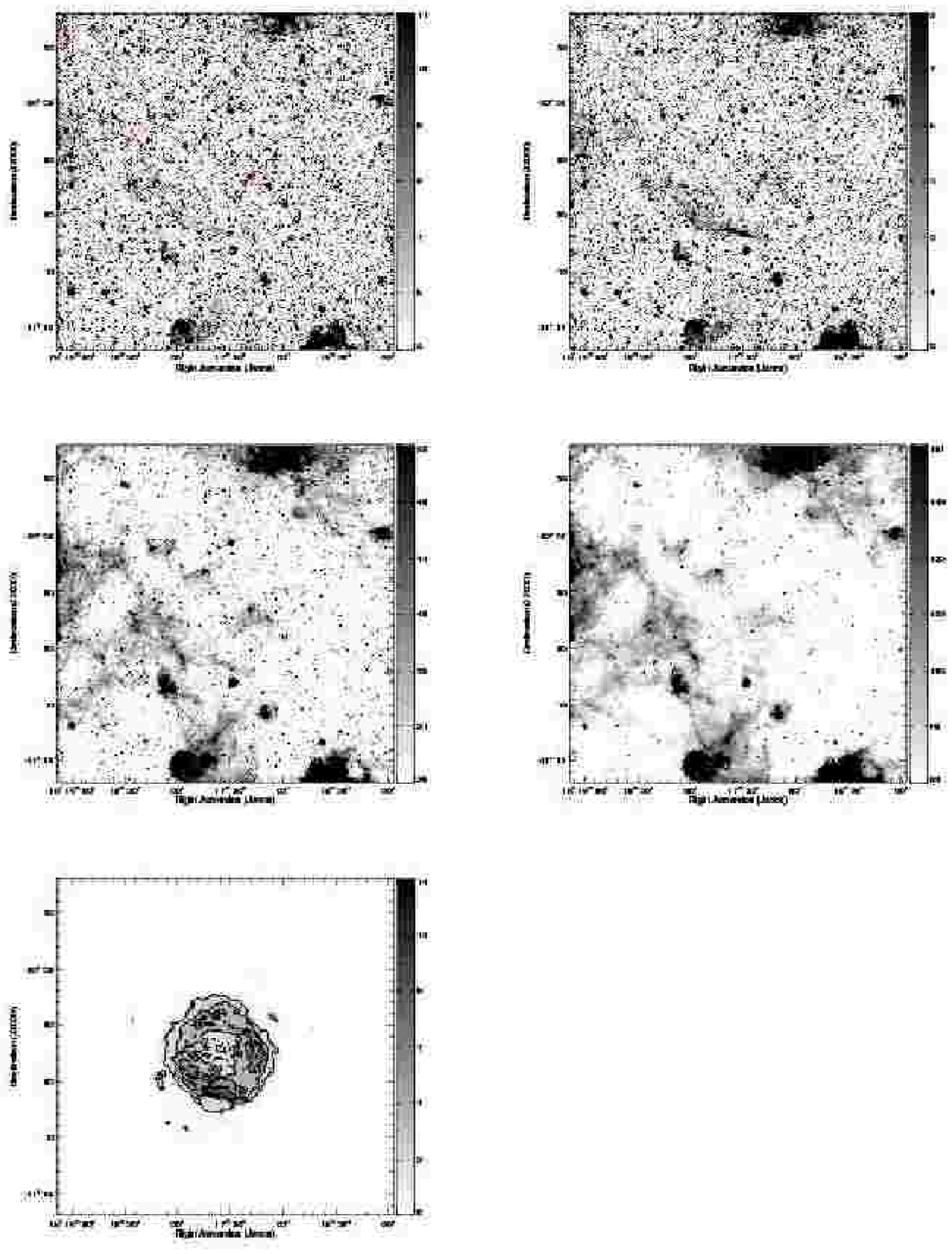}}
{\bf Fig. 6.}---~
IRAC images of G332.4--0.4 (RCW 103) at
   3.6 \micron\ (top-left),
   4.5 \micron\ (top-right),
   5.8 \micron\ (middle-left), and
   8.0 \micron\ (middle-right).
   The gray-scale range
   is shown at the left of each image.
   ATCA 2.32 GHz continuum image with
   $4^{\prime \prime}.4 \! \times \! 3^{\prime \prime}.7$ beam (Dickel et al. 1996)
   is also shown at the bottom-left.
   The contour levels are
   1, 3, 5, and 7 mJy beam$^{-1}$.
\end{figure*}

\newpage
\begin{figure*}
\vspace{-3cm}
\centerline{\epsfysize=24cm\epsfbox{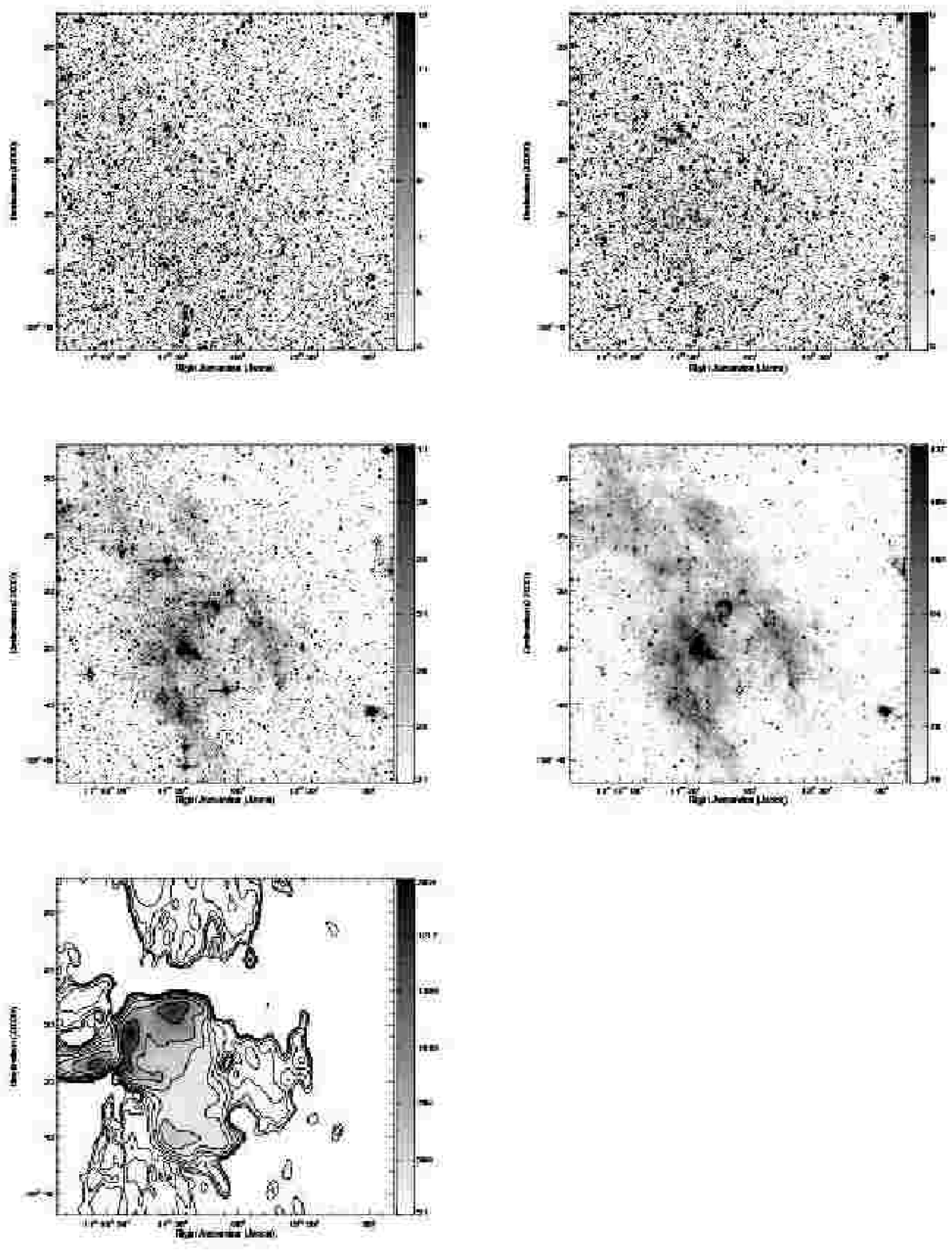}}
{\bf Fig. 7.}---~
IRAC images of G348.5+0.1 (CTB 37A) at
   3.6 \micron\ (top-left),
   4.5 \micron\ (top-right),
   5.8 \micron\ (middle-left), and
   8.0 \micron\ (middle-right).
   The gray-scale range
   is shown at the left of each image.
   MOST 0.843 GHz continuum image with
   $43^{\prime \prime} \! \times \! 43^{\prime \prime}$ beam
   is also shown at the bottom-left.
   The contour levels are
   5, 10, 20, 40, 80, 160, 320, 640, and 1280 mJy beam$^{-1}$.
\end{figure*}

\newpage
\begin{figure*}
\vspace{-3cm}
\centerline{\epsfysize=24cm\epsfbox{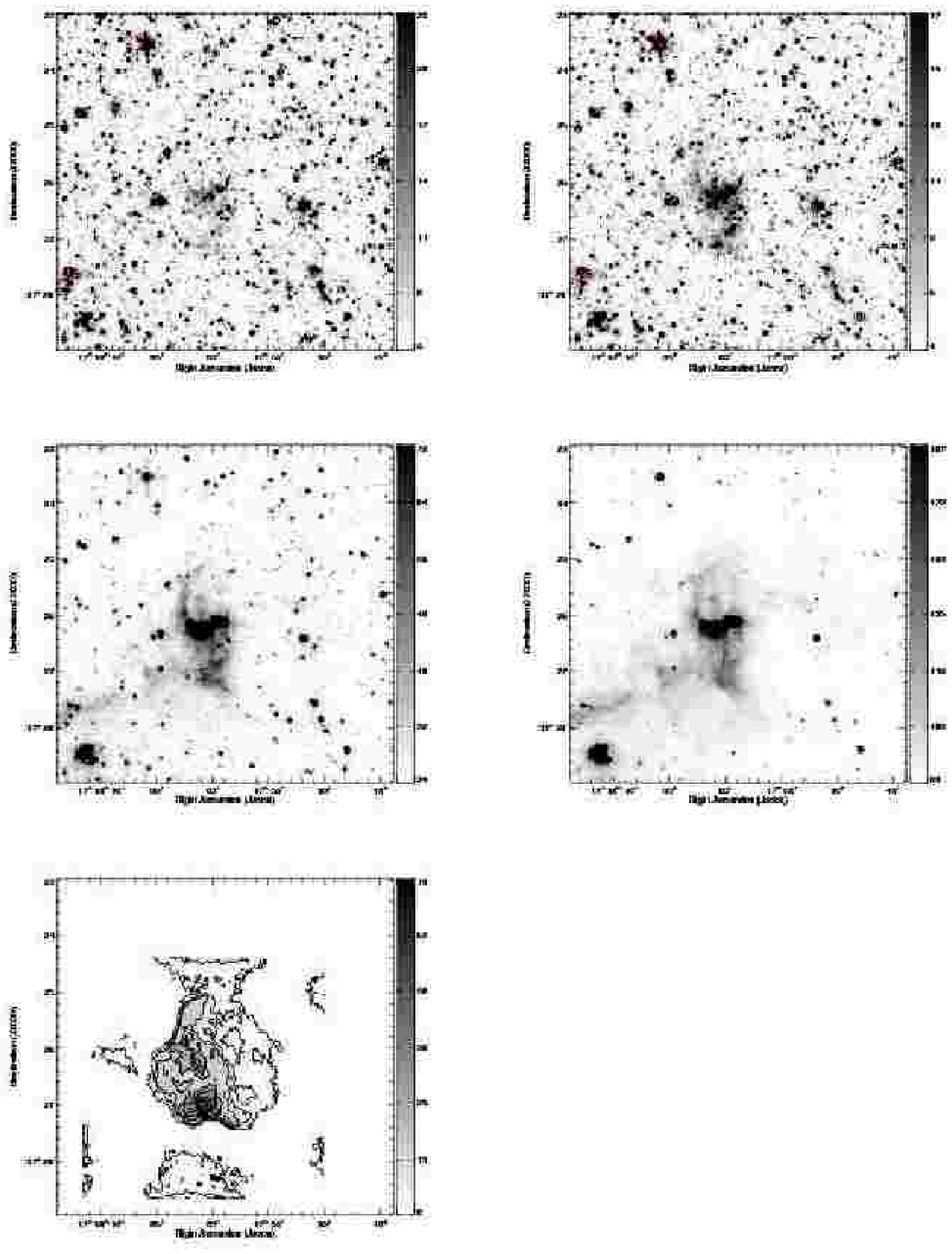}}
{\bf Fig. 8.}---~
IRAC images of G349.7+0.2 at
   3.6 \micron\ (top-left),
   4.5 \micron\ (top-right),
   5.8 \micron\ (middle-left), and
   8.0 \micron\ (middle-right).
   The gray-scale range
   is shown at the left of each image.
   VLA 1.51 GHz continuum image with
   $5^{\prime \prime}.0 \! \times \! 2^{\prime \prime}.1$ beam (Brogan et al. 2000)
   is also shown at the bottom-left.
   The contour levels are
   1, 5, 10, 20, 30, 40, and 50 mJy beam$^{-1}$.
\end{figure*}

\newpage
\begin{figure*}
\vspace{-3cm}
\centerline{\epsfysize=16cm\epsfbox{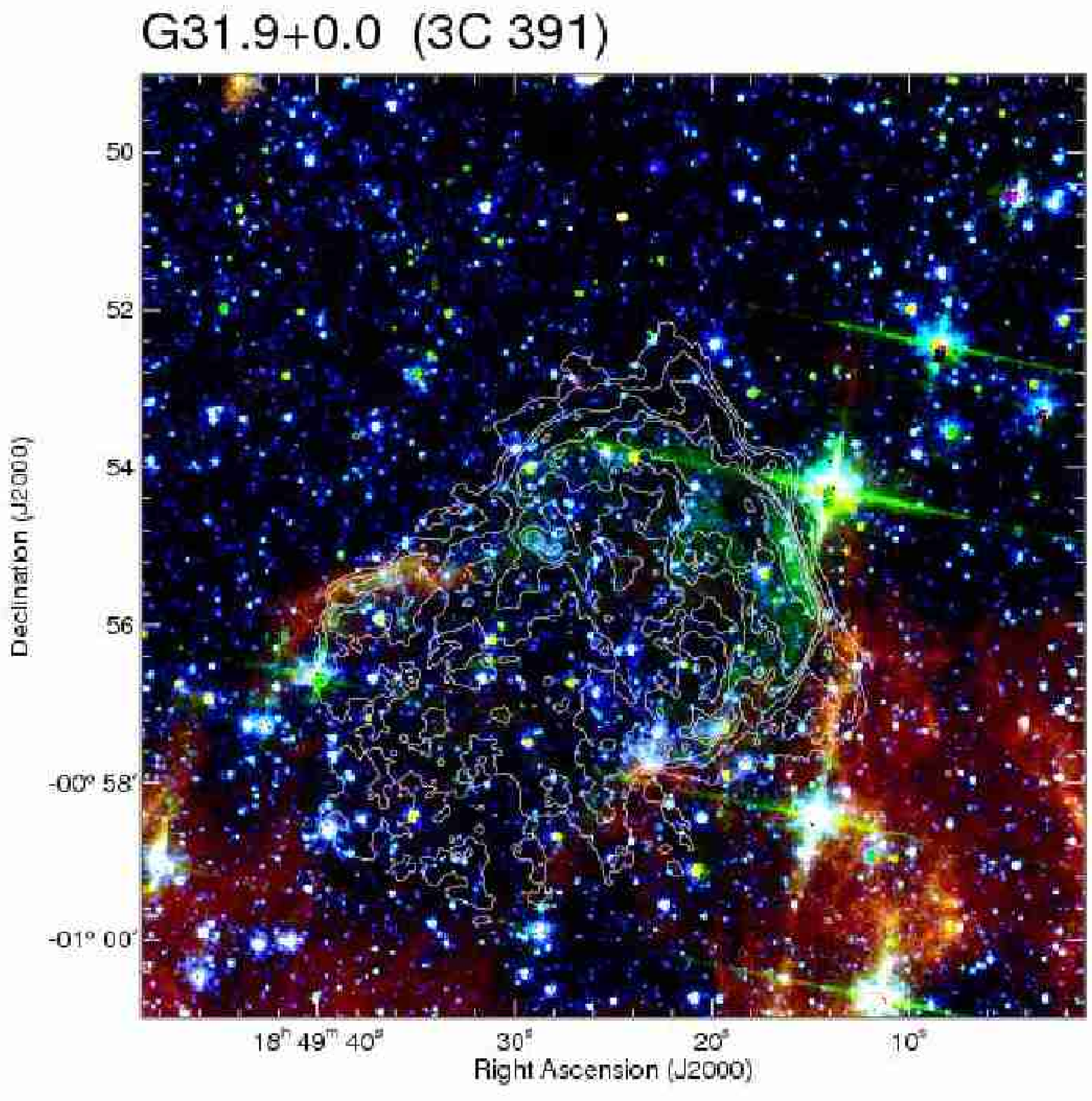}}
{\bf Fig. 9.}---~
G31.9+0.0 (3C 391)
   R (8.0 \micron),  G (5.8 \micron), B (4.5 \micron) color image
   with VLA 1.46 GHz continuum map superposed.
   Color ranges and contour levels are the same as Fig.~1.
\end{figure*}

\newpage
\begin{figure*}
\vspace{-3cm}
\centerline{\epsfysize=16cm\epsfbox{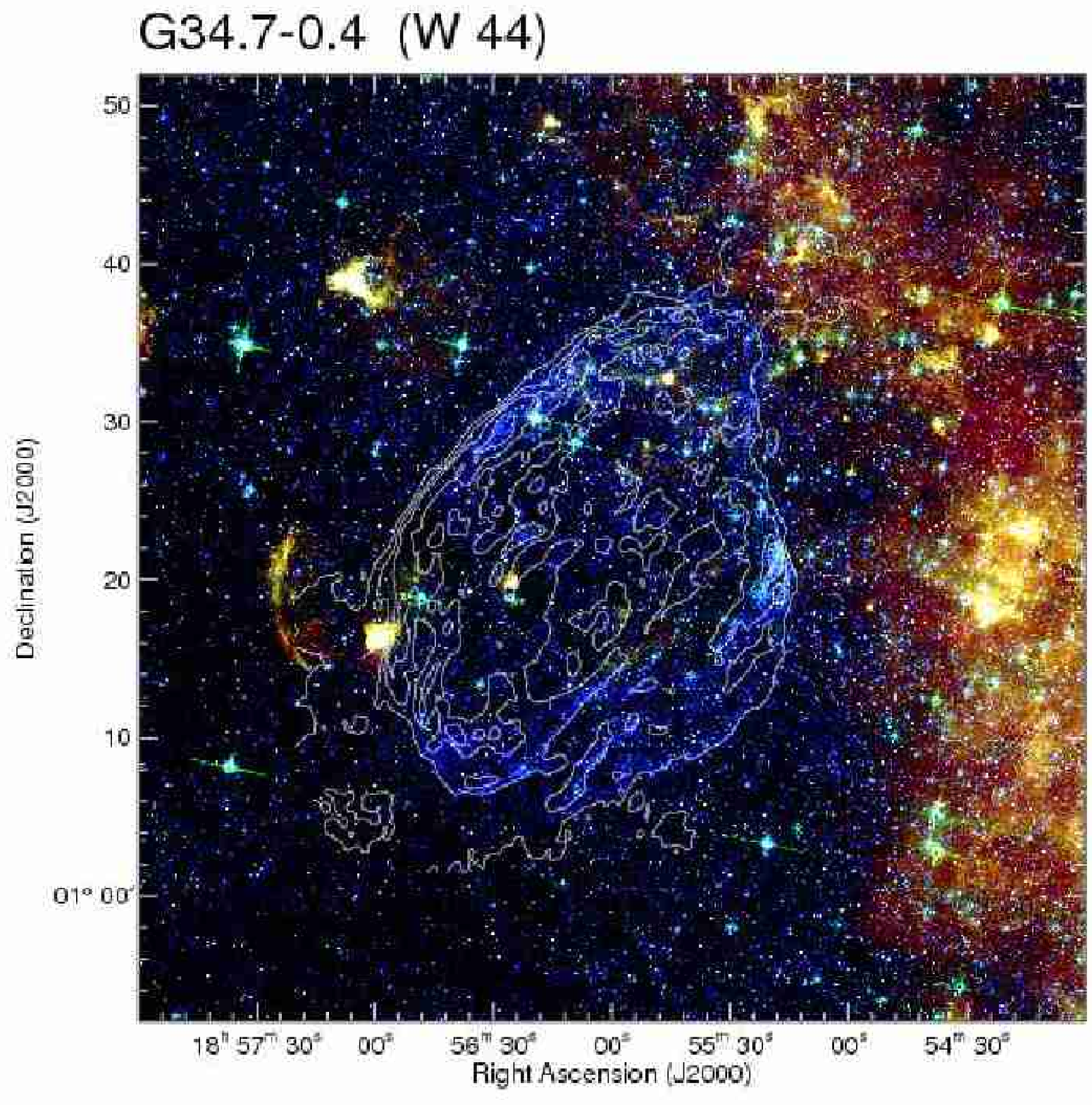}}
{\bf Fig. 10.}---~
G34.7--0.4 (W 44)
   R (8.0 \micron),  G (5.8 \micron), B (4.5 \micron) color image
   with VLA 1.44 GHz continuum map superposed.
   Color ranges and contour levels are the same as Fig.~2.
\end{figure*}

\newpage
\begin{figure*}
\vspace{-3cm}
\centerline{\epsfysize=16cm\epsfbox{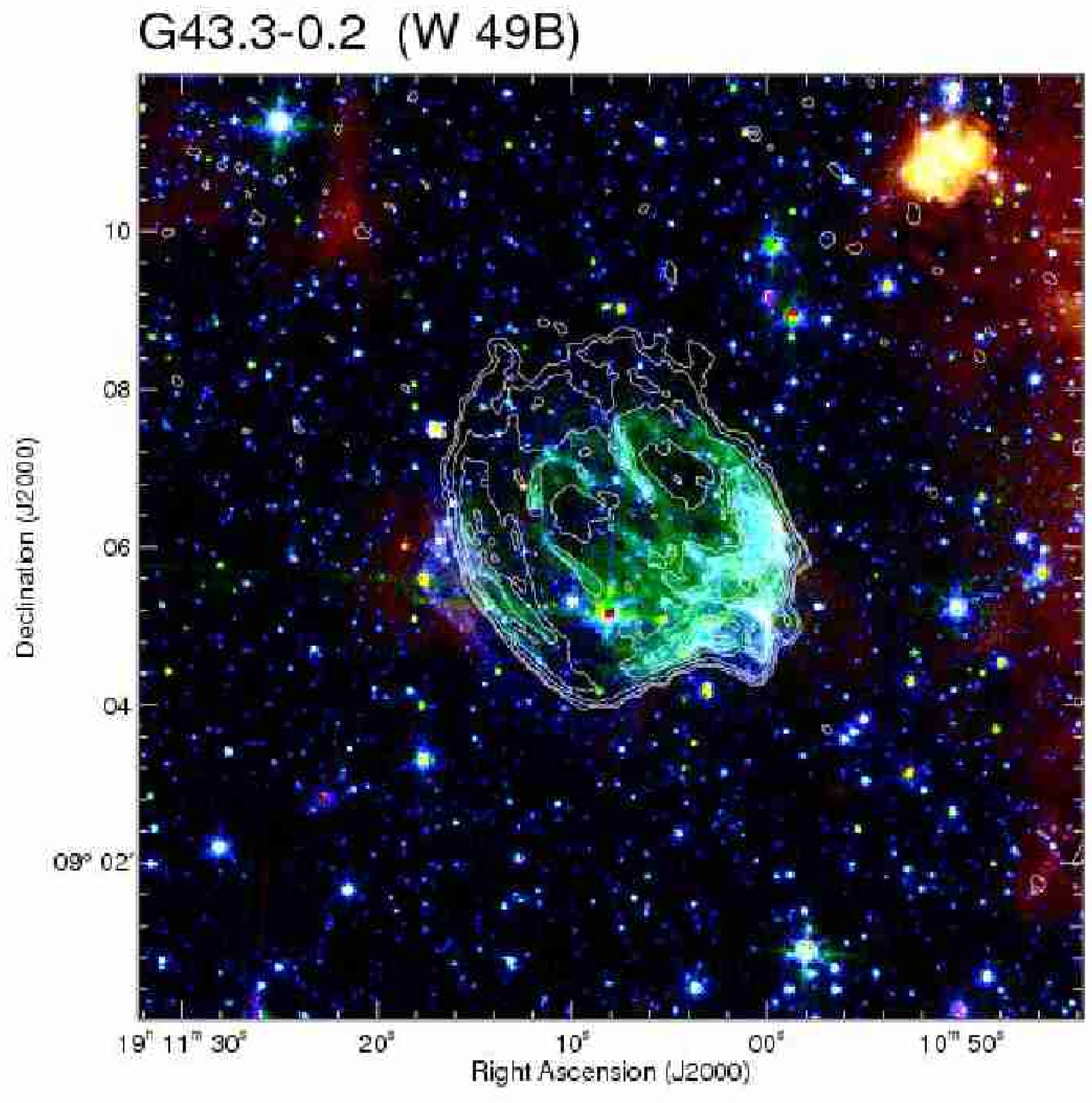}}
{\bf Fig. 11.}---~
G43.3--0.2l(W 49B)
   R (8.0 \micron),  G (5.8 \micron), B (4.5 \micron) color image
   with VLA 1.45 GHz continuum map superposed.
   Color ranges and contour levels are the same as Fig.~3.
\end{figure*}

\newpage
\begin{figure*}
\vspace{-3cm}
\centerline{\epsfysize=16cm\epsfbox{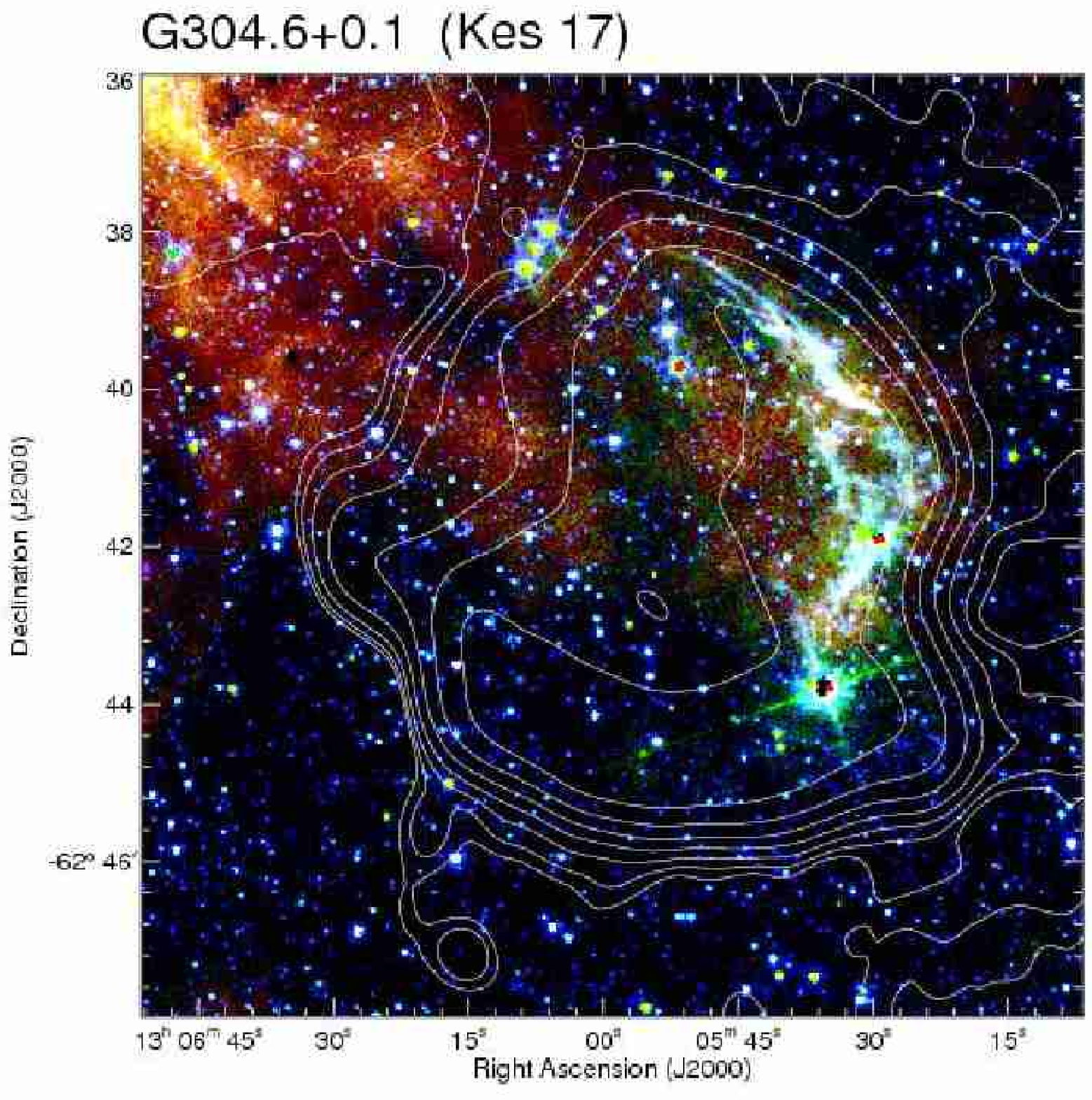}}
{\bf Fig. 12.}---~
G304.6+0.1 (Kes 17)
   R (8.0 \micron),  G (5.8 \micron), B (4.5 \micron) color image
   with MOST 0.843 GHz continuum map superposed.
   Color ranges and contour levels are the same as Fig.~4.
\end{figure*}

\newpage
\begin{figure*}
\vspace{-3cm}
\centerline{\epsfysize=16cm\epsfbox{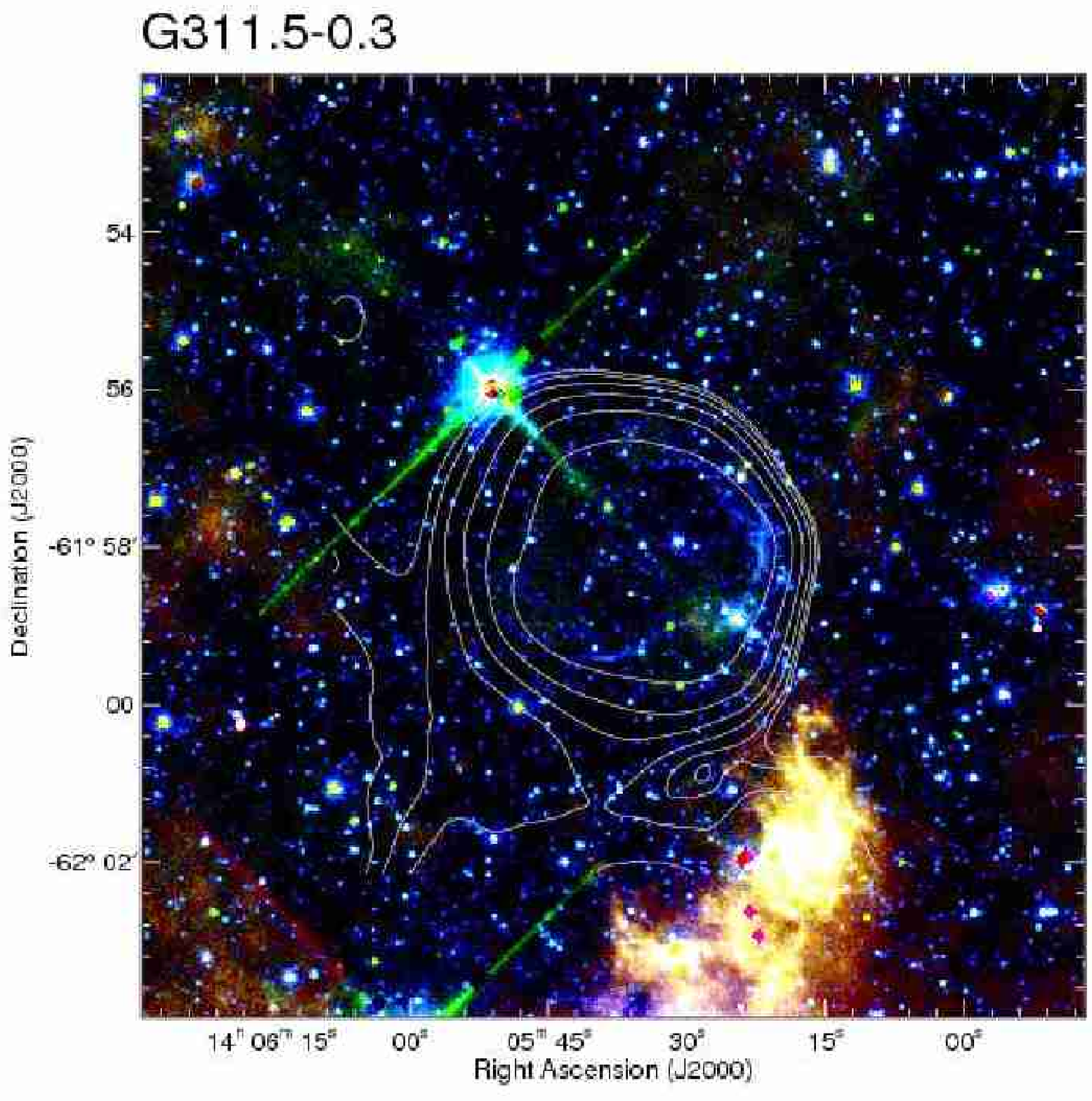}}
{\bf Fig. 13.}---~
G311.5--0.3
   R (8.0 \micron),  G (5.8 \micron), B (4.5 \micron) color image
   with MOST 0.843 GHz continuum map superposed.
   Color ranges and contour levels are the same as Fig.~5.
\end{figure*}

\newpage
\begin{figure*}
\vspace{-3cm}
\centerline{\epsfysize=16cm\epsfbox{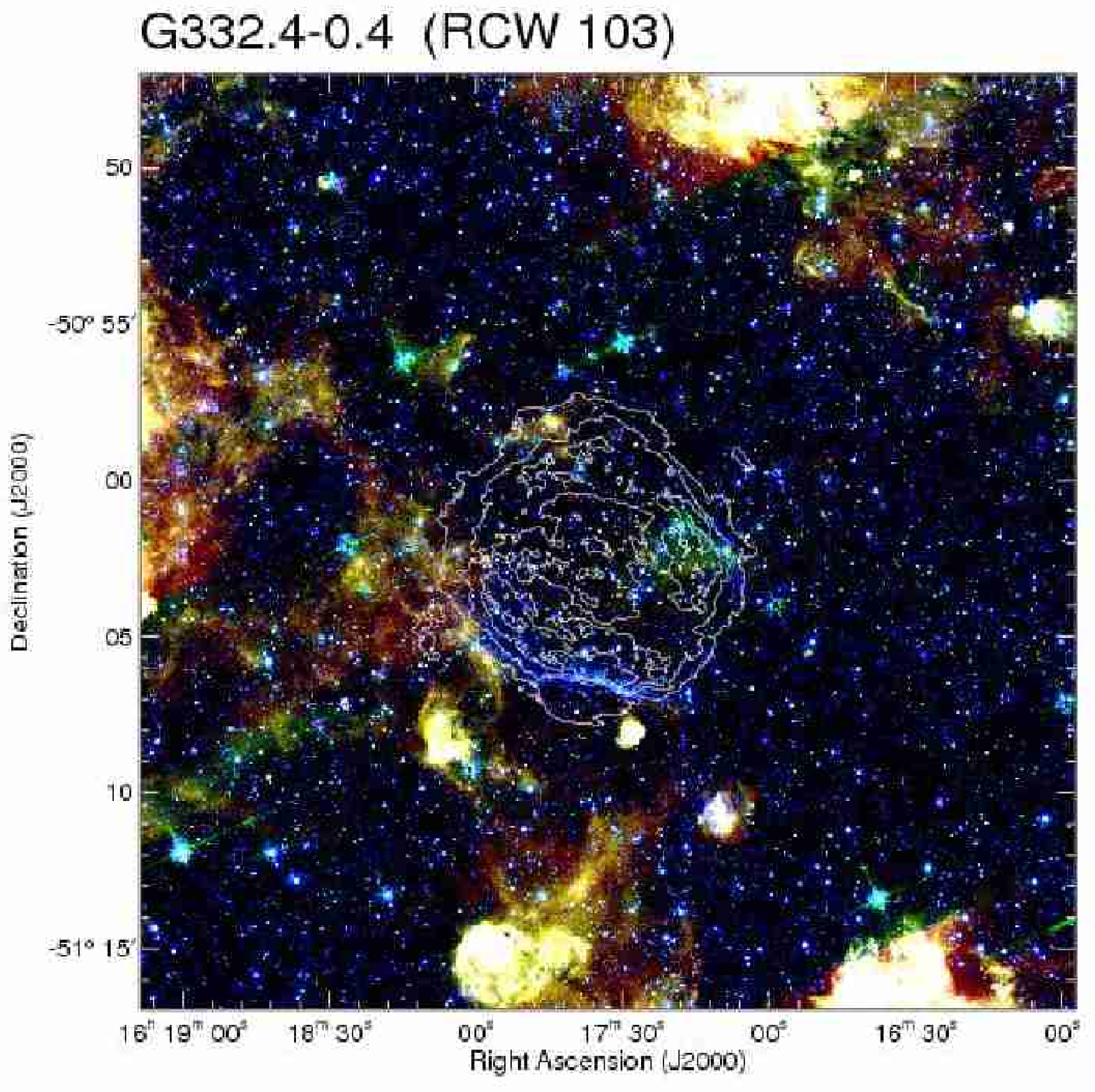}}
{\bf Fig. 14.}---~
G332.4--0.4 (RCW 103)
   R (8.0 \micron),  G (5.8 \micron), B (4.5 \micron) color image
   with ATCA 2.32 GHz continuum map superposed.
   Color ranges and contour levels are the same as Fig.~6.
\end{figure*}

\newpage
\begin{figure*}
\vspace{-3cm}
\centerline{\epsfysize=16cm\epsfbox{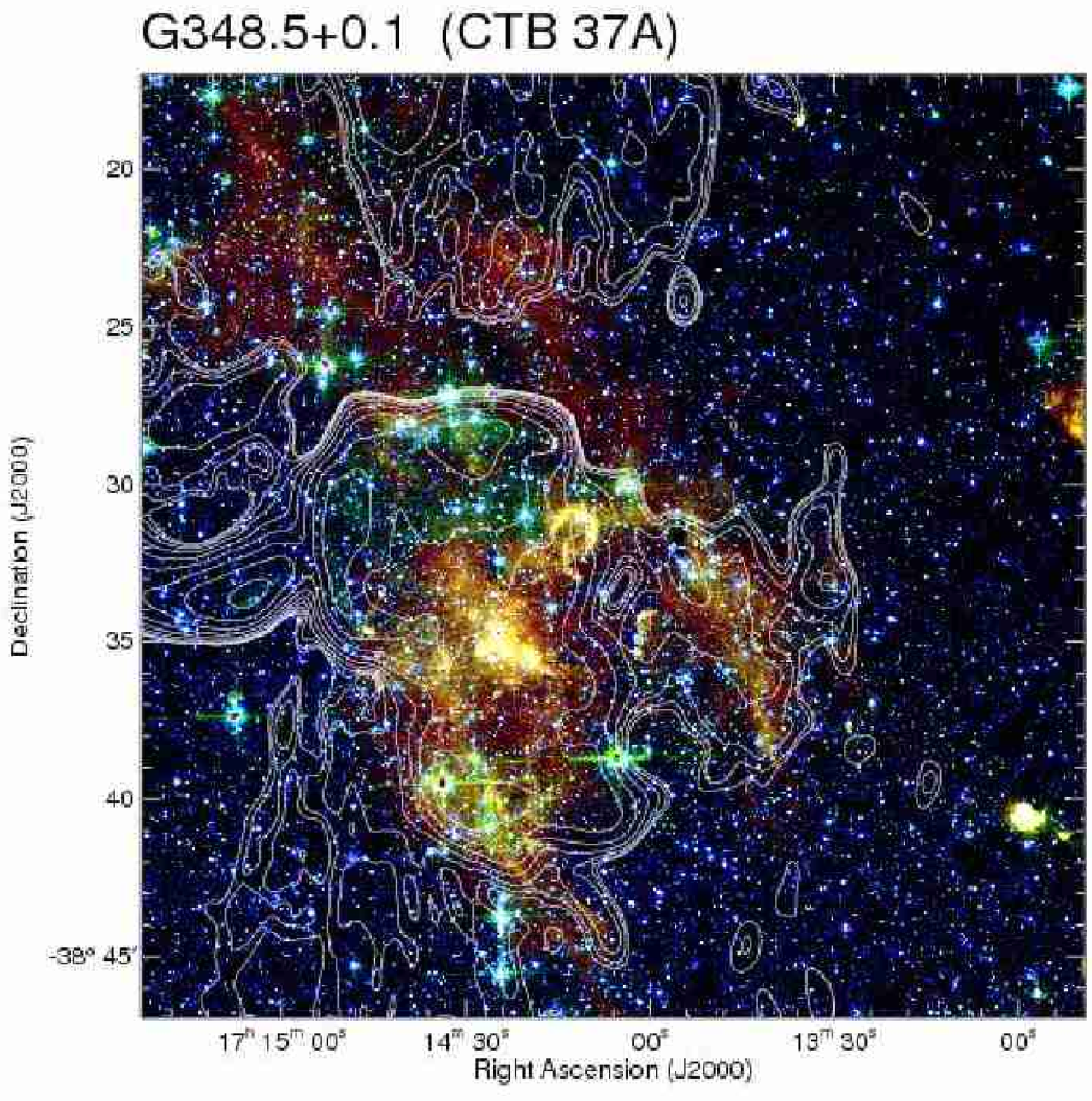}}
{\bf Fig. 15.}---~
G348.5+0.1 (CTB 37A)
   R (8.0 \micron),  G (5.8 \micron), B (4.5 \micron) color image
   with MOST 0.843 GHz continuum map superposed.
   Color ranges and contour levels are the same as Fig.~7.
\end{figure*}

\newpage
\begin{figure*}
\vspace{-3cm}
\centerline{\epsfysize=16cm\epsfbox{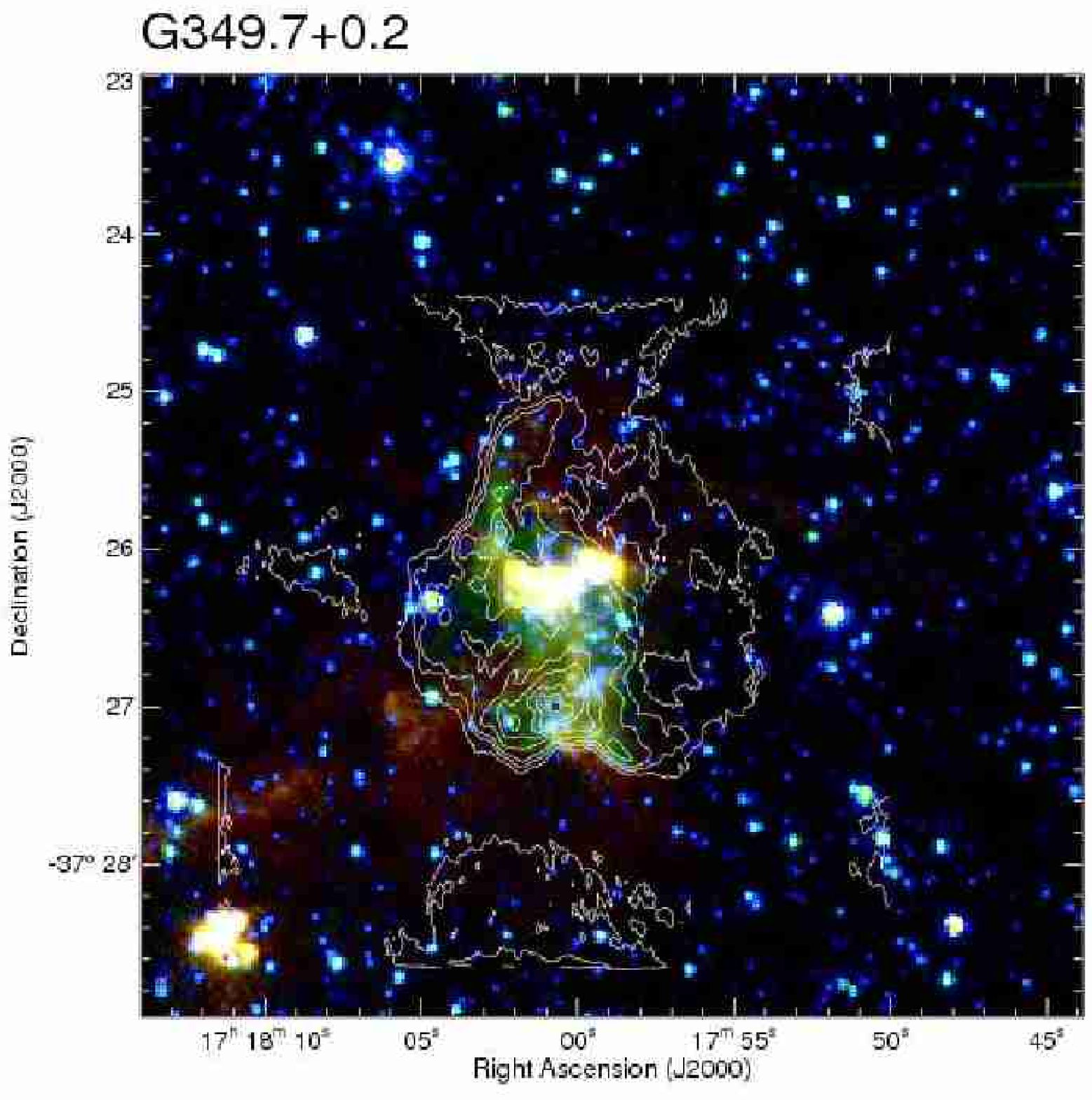}}
{\bf Fig. 16.}---~
G349.7+0.2
   R (8.0 \micron),  G (5.8 \micron), B (4.5 \micron) color image
   with VLA 1.51 GHz continuum map superposed.
   Color ranges and contour levels are the same as Fig.~8.
\end{figure*}

\newpage
\begin{figure*}
\vspace{-3cm}
\centerline{\epsfysize=24cm\epsfbox{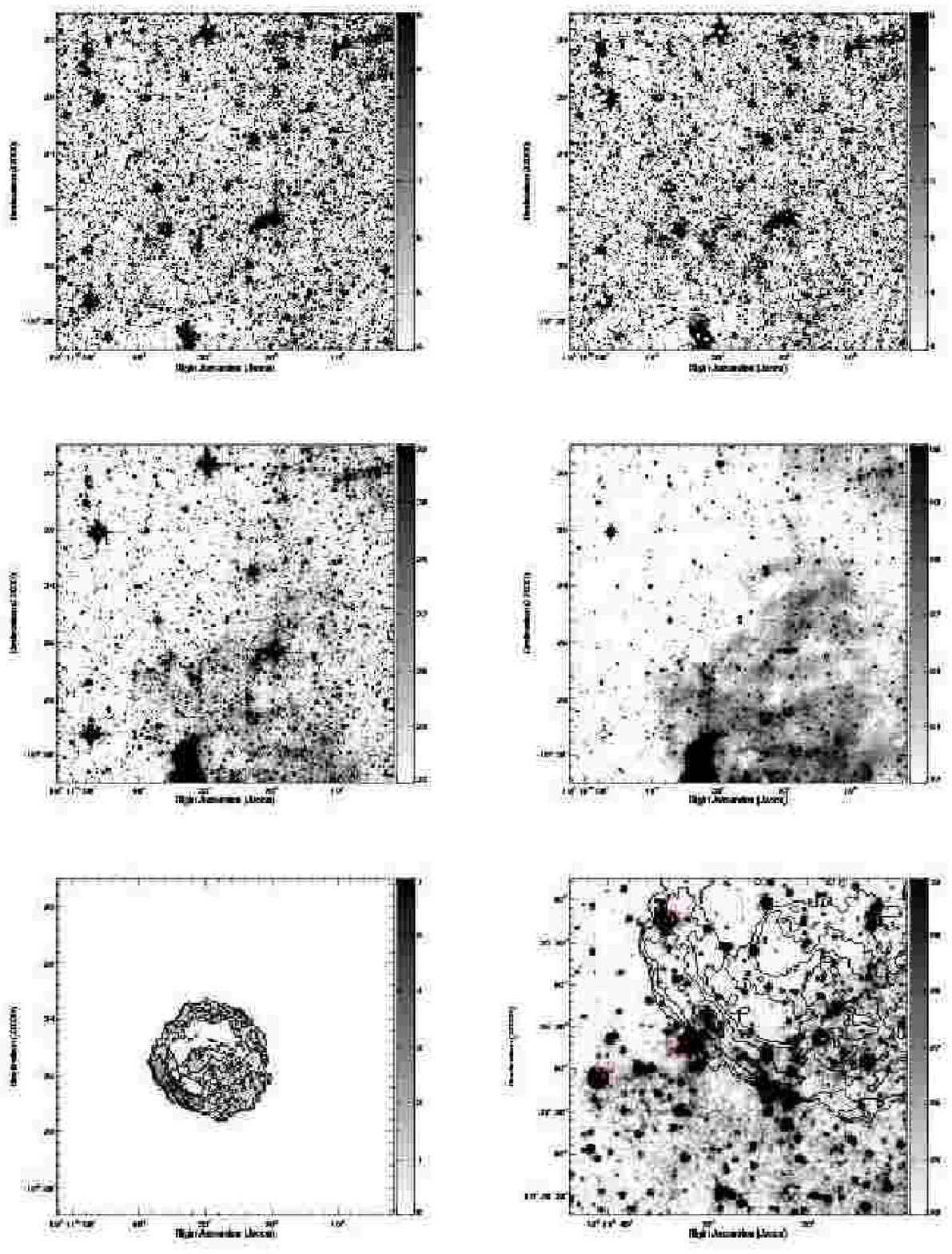}}
{\bf Fig. 17.}---~
IRAC images of G11.2--0.3 at
   3.6 \micron\ (top-left),
   4.5 \micron\ (top-right),
   5.8 \micron\ (middle-left), and
   8.0 \micron\ (middle-right).
   The gray-scale range
   is shown at the left of each image.
   VLA 4.76 GHz continuum image with
   $3^{\prime \prime}.0 \! \times \! 3^{\prime \prime}.0$ beam (Green et al. 1988)
   is also shown at the bottom-left.
   The contour levels are
   50 ,100, 200, and 400 mJy beam$^{-1}$.
   IRAC 5.8 \micron\ zoom up image of infrared emission
   with VLA 4.76 GHz continuum contour
   is shown at bottom-right.
\end{figure*}

\newpage
\begin{figure*}
\vspace{-3cm}
\centerline{\epsfysize=24cm\epsfbox{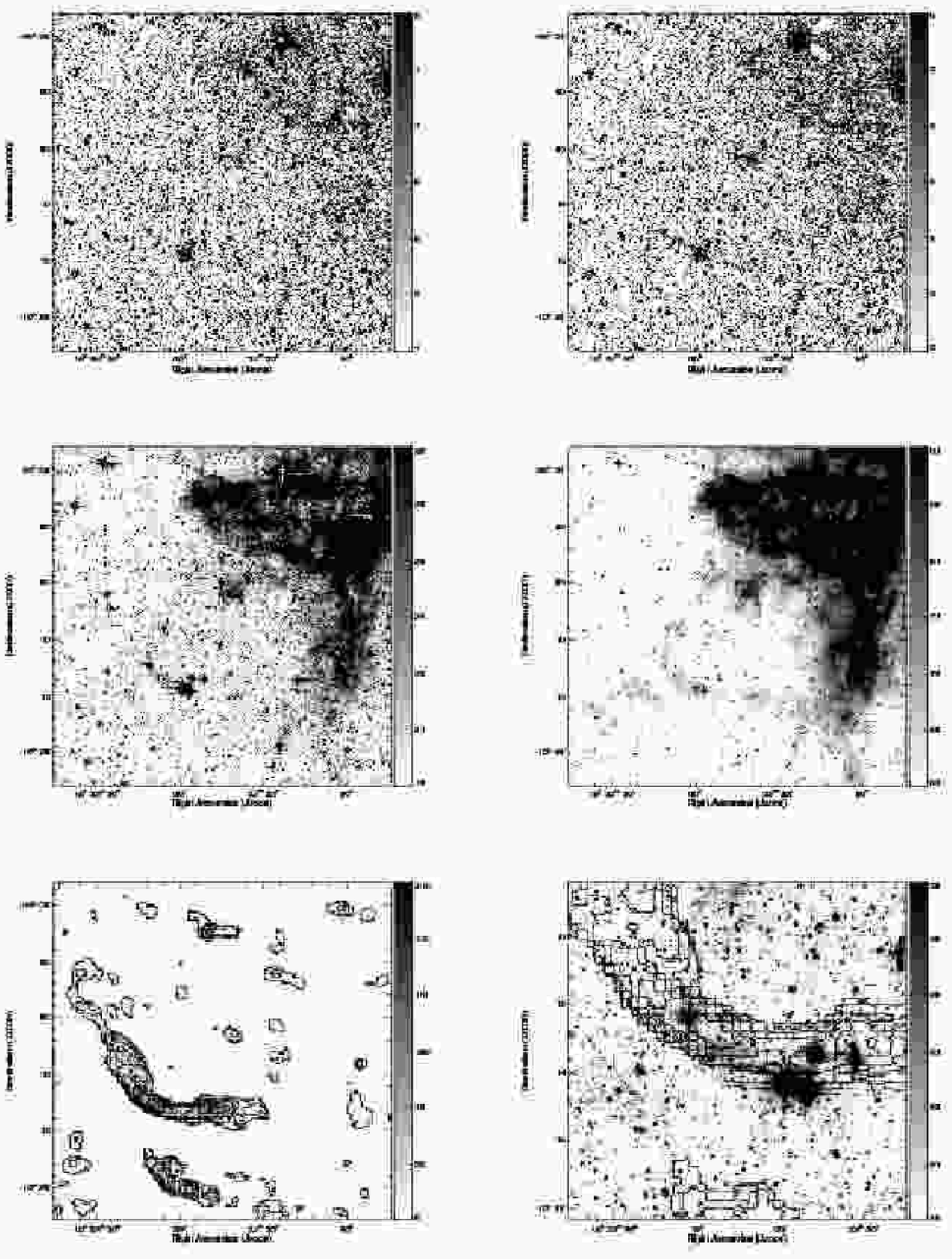}}
{\bf Fig. 18.}---~
IRAC images of G21.8--0.6 (Kes 69) at
   3.6 \micron\ (top-left),
   4.5 \micron\ (top-right),
   5.8 \micron\ (middle-left), and
   8.0 \micron\ (middle-right).
   The gray-scale range
   is shown at the left of each image.
   NVSS 1.4 GHz continuum image with
   $45^{\prime \prime}.0 \! \times \! 45^{\prime \prime}.0$ beam
   is also shown at the bottom-left.
   The contour levels are
   5, 10, 20, 40, 80, and 160 mJy beam$^{-1}$.
   IRAC 5.8 \micron\ zoom up image of infrared emission
   with NVSS 1.4 GHz continuum contour
   is shown at bottom-right.
\end{figure*}

\clearpage
\newpage
\begin{figure*}
\vspace{-3cm}
\centerline{\epsfysize=24cm\epsfbox{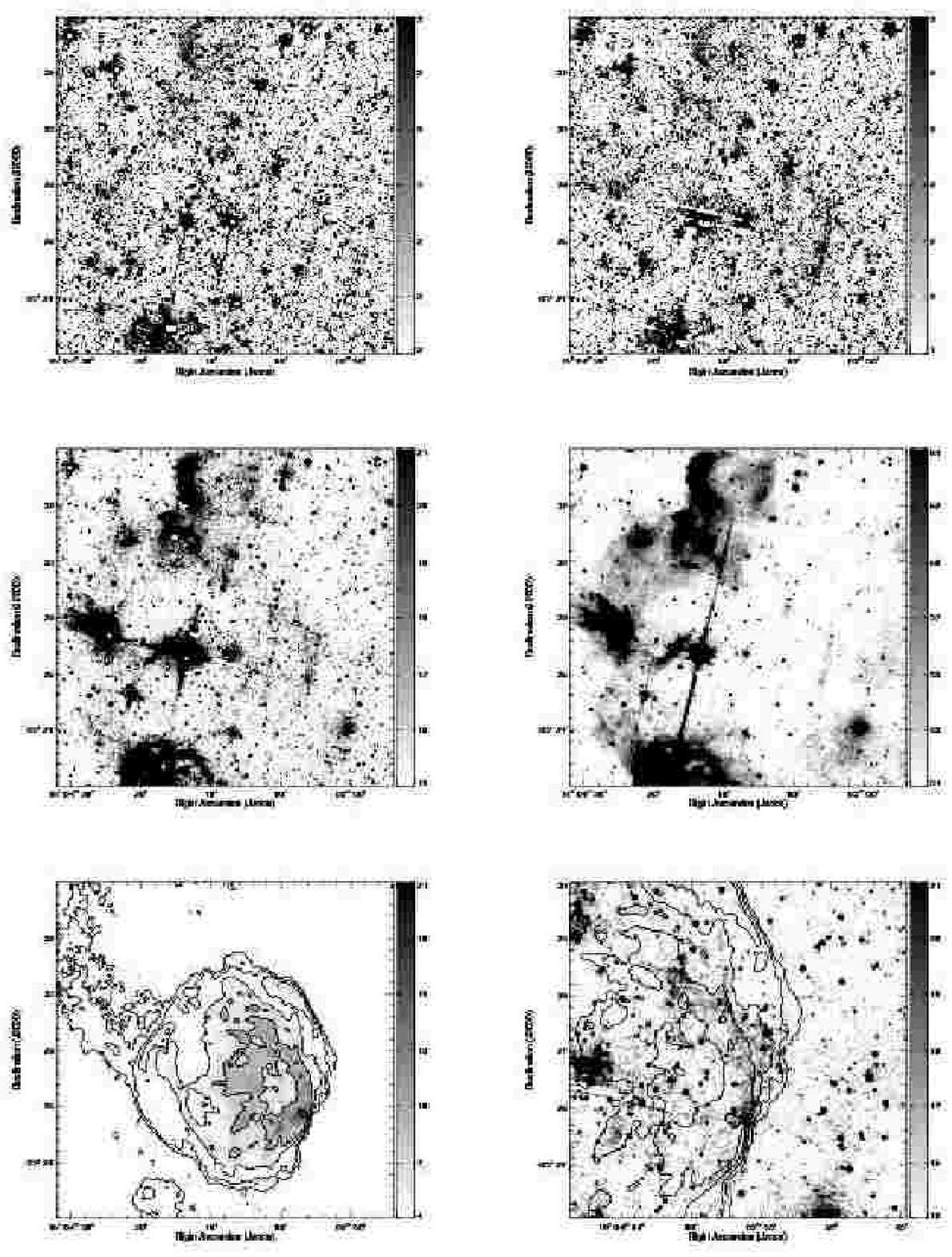}}
{\bf Fig. 19.}---~
IRAC images of G39.2--0.3 (3C 396) at
   3.6 \micron\ (top-left),
   4.5 \micron\ (top-right),
   5.8 \micron\ (middle-left), and
   8.0 \micron\ (middle-right).
   The gray-scale range
   is shown at the left of each image.
   VLA 1.47 GHz continuum image with
   $6^{\prime \prime}.8 \! \times \! 6^{\prime \prime}.1$ beam (Dyer \& Reynolds (1999a)
   is also shown at the bottom-left.
   The contour levels are
   1, 2, 4, 8, and 16 mJy beam$^{-1}$.
   IRAC 5.8 \micron\ zoom up image of infrared emission
   with VLA 1.47 GHz continuum contour
   is shown at bottom-right.
\end{figure*}

\newpage
\begin{figure*}
\vspace{-3cm}
\centerline{\epsfysize=24cm\epsfbox{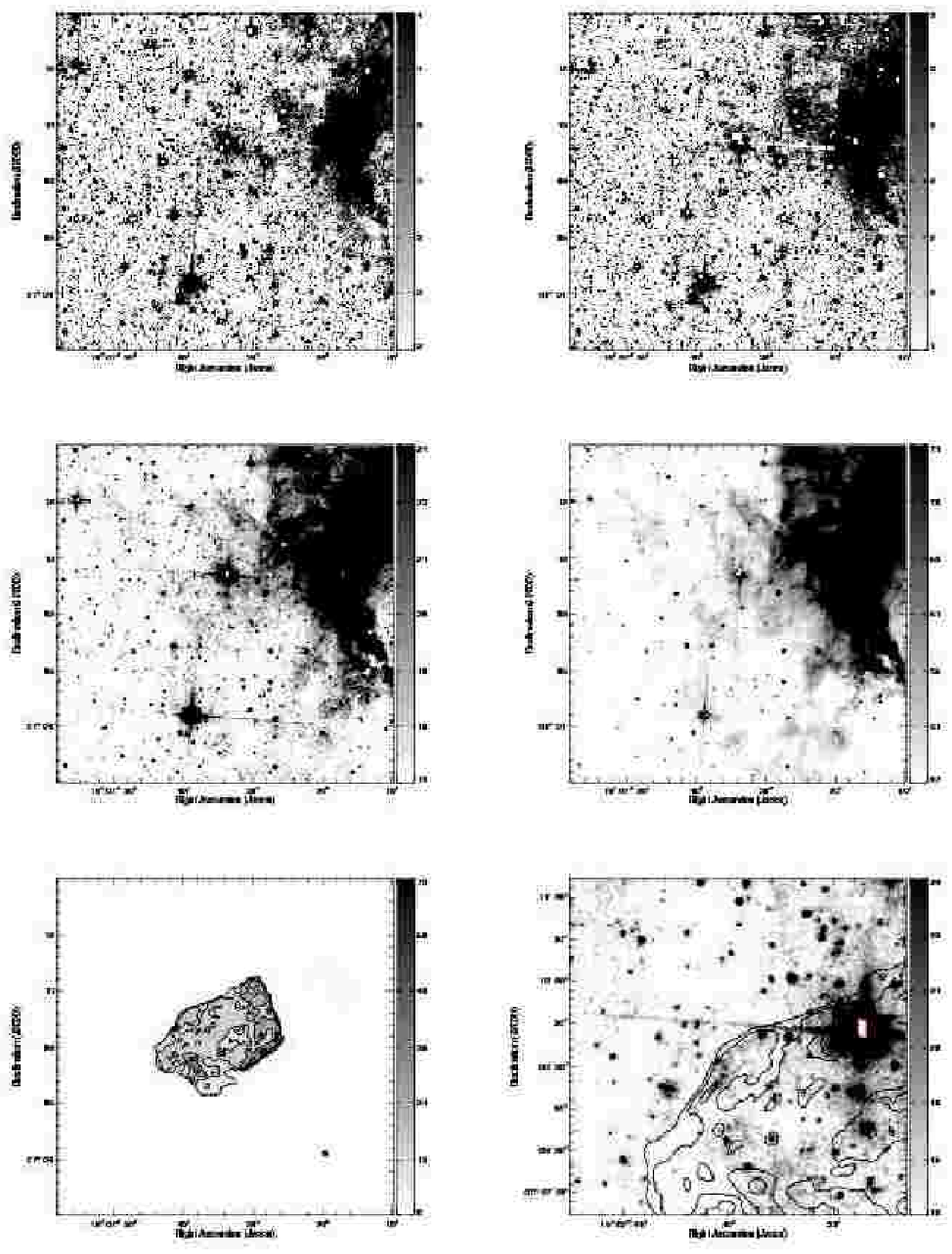}}
{\bf Fig. 20.}---~
IRAC images of G41.1--0.3 (3C 397) at
   3.6 \micron\ (top-left),
   4.5 \micron\ (top-right),
   5.8 \micron\ (middle-left), and
   8.0 \micron\ (middle-right).
   The gray-scale range
   is shown at the left of each image.
   VLA 1.47 GHz continuum image with
   $6^{\prime \prime}.4 \! \times \! 6^{\prime \prime}.1$ beam (Dyer \& Reynolds (1999b)
   is also shown at the bottom-left.
   The contour levels are
   5, 10, 20, 40, and 80 mJy beam$^{-1}$.
   IRAC 5.8 \micron\ zoom up image of infrared emission
   with VLA 1.47 GHz continuum contour
   is shown at bottom-right.
\end{figure*}

\newpage
\begin{figure*}
\vspace{-3cm}
\centerline{\epsfysize=24cm\epsfbox{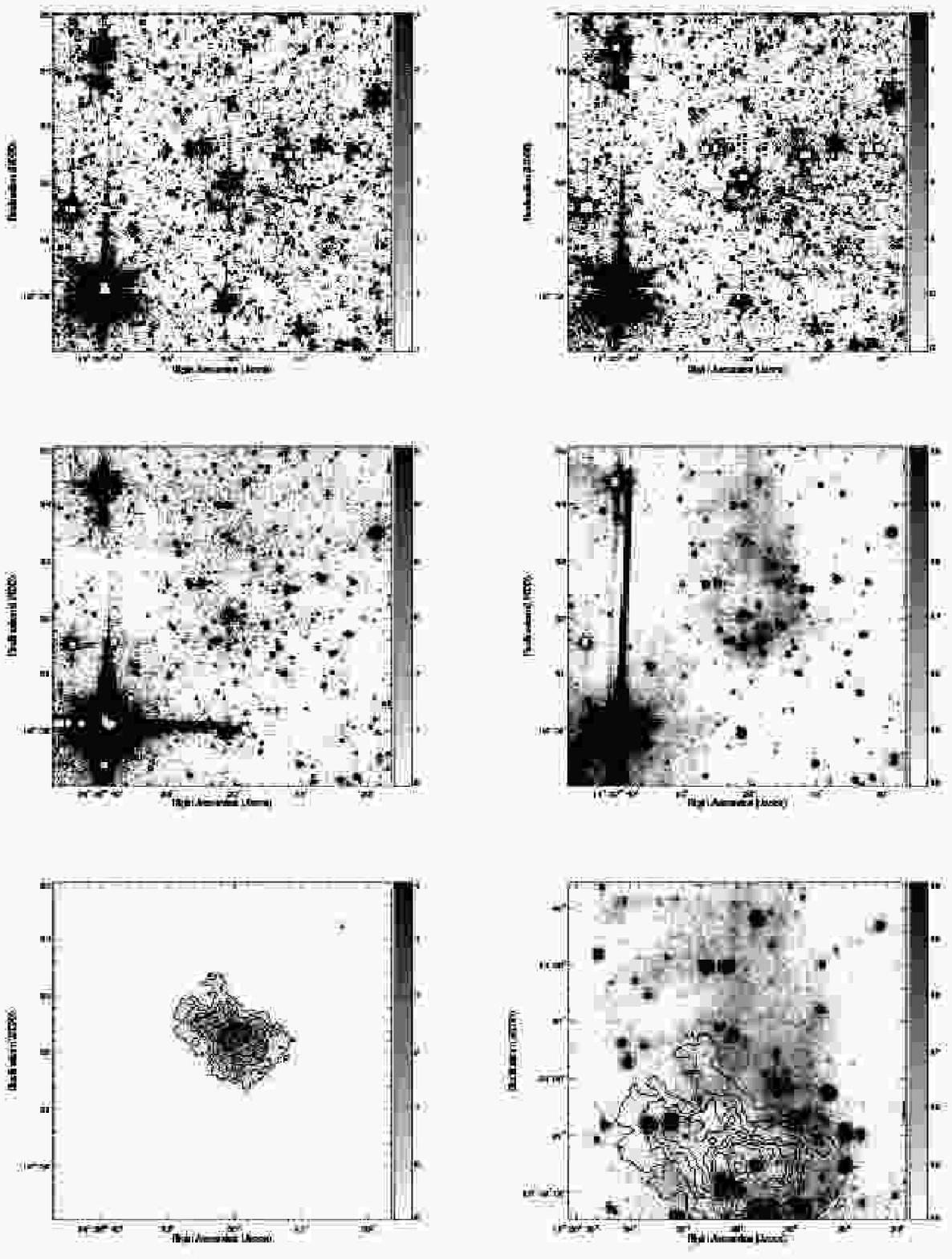}}
{\bf Fig. 21.}---~
IRAC images of G54.1+0.3 at
   3.6 \micron\ (top-left),
   4.5 \micron\ (top-right),
   5.8 \micron\ (middle-left), and
   8.0 \micron\ (middle-right).
   The gray-scale range
   is shown at the left of each image.
   VLA 4.85 GHz continuum image with
   $6^{\prime \prime}.2 \! \times \! 5^{\prime \prime}.2$ beam
   is also shown at the bottom-left.
   The contour levels are
   5, 10, 20, 40, 80, and 160 mJy beam$^{-1}$.
   IRAC 8.0 \micron\ zoom up image of infrared emission
   with VLA 4.85 GHz continuum contour
   is shown at bottom-right.
\end{figure*}

\newpage
\begin{figure*}
\vspace{-3cm}
\centerline{\epsfysize=24cm\epsfbox{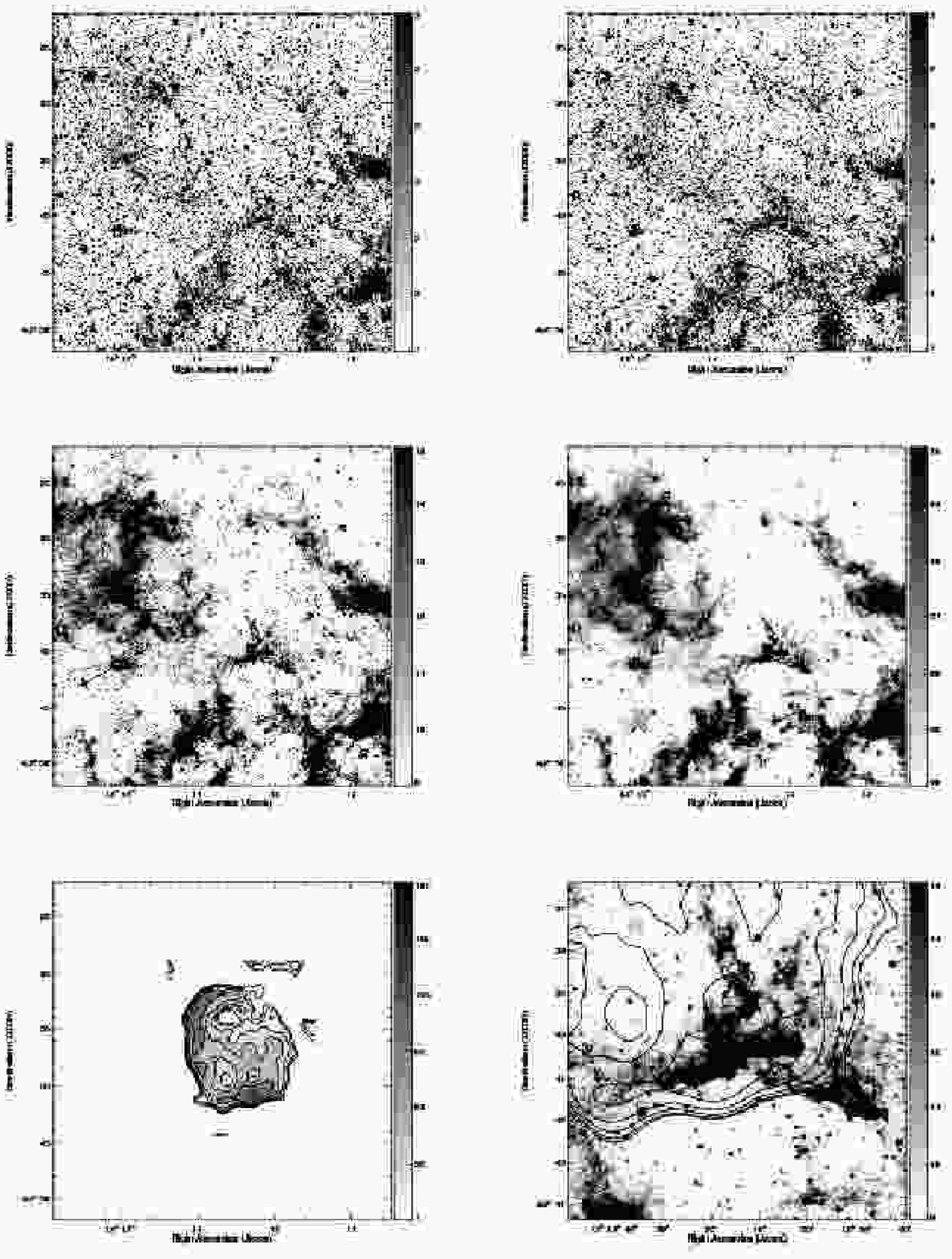}}
{\bf Fig. 22.}---~
IRAC images of G298.6--0.0 at
   3.6 \micron\ (top-left),
   4.5 \micron\ (top-right),
   5.8 \micron\ (middle-left), and
   8.0 \micron\ (middle-right).
   The gray-scale range
   is shown at the left of each image.
   MOST 0.843 GHz continuum image with
   $43^{\prime \prime} \! \times \! 43^{\prime \prime}$ beam
   is also shown at the bottom-left.
   The contour levels are
   5, 10, 20, 40, 80, and 160 mJy beam$^{-1}$.
   IRAC 5.8 \micron\ zoom up image of infrared emission
   with MOST 0.843 GHz continuum contour
   is shown at bottom-right.
\end{figure*}

\newpage
\begin{figure*}
\vspace{-3cm}
\centerline{\epsfysize=24cm\epsfbox{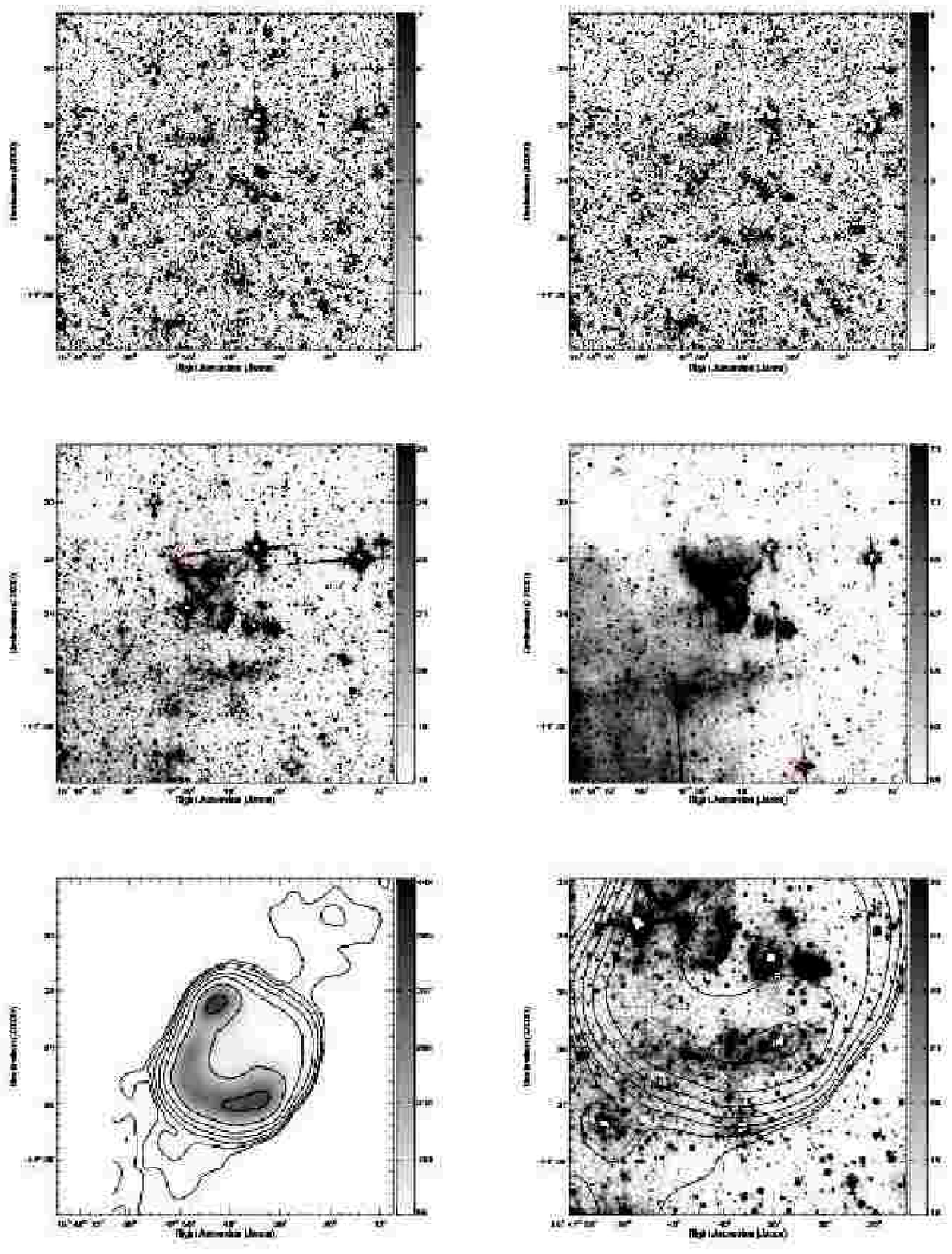}}
{\bf Fig. 23.}---~
IRAC images of G340.6+0.3 at
   3.6 \micron\ (top-left),
   4.5 \micron\ (top-right),
   5.8 \micron\ (middle-left), and
   8.0 \micron\ (middle-right).
   The gray-scale range
   is shown at the left of each image.
   MOST 0.843 GHz continuum image with
   $43^{\prime \prime} \! \times \! 43^{\prime \prime}$ beam
   is also shown at the bottom-left.
   The contour levels are
   5, 10, 20, 40, 80, 160, and 320 mJy beam$^{-1}$.
   IRAC 5.8 \micron\ zoom up image of infrared emission
   with MOST 0.843 GHz continuum contour
   is shown at bottom-right.
\end{figure*}

\newpage
\begin{figure*}
\vspace{-3cm}
\centerline{\epsfysize=24cm\epsfbox{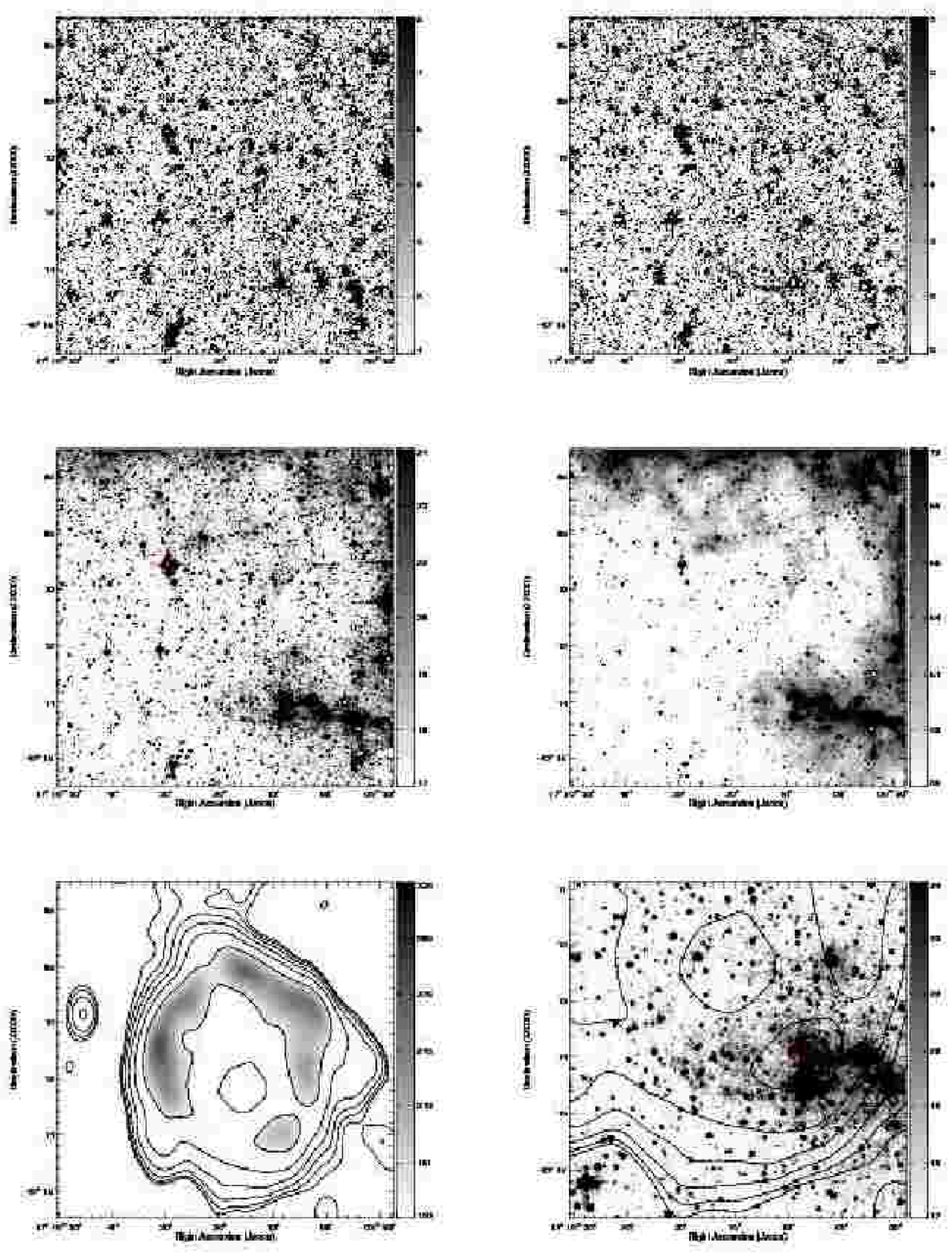}}
{\bf Fig. 24.}---~
IRAC images of G346.6--0.2 at
   3.6 \micron\ (top-left),
   4.5 \micron\ (top-right),
   5.8 \micron\ (middle-left), and
   8.0 \micron\ (middle-right).
   The gray-scale range
   is shown at the left of each image.
   MOST 0.843 GHz continuum image with
   $43^{\prime \prime} \! \times \! 43^{\prime \prime}$ beam
   is also shown at the bottom-left.
   The contour levels are
   5, 10, 20, 40, 80, and 160 mJy beam$^{-1}$.
   IRAC 5.8 \micron\ zoom up image of infrared emission
   with MOST 0.843 GHz continuum contour
   is shown at bottom-right.
\end{figure*}

\end{document}